\documentclass[prd,showpacs,showkeys,nofootinbib,12pt%,twocolumn
]{revtex4}

\usepackage{epsfig}
\usepackage{amssymb}

\usepackage{latexsym}
\def\qslash{\not\!Q}
\def\zslash{\not\!Z}

\def\a{\alpha}
\def\b{\beta}
\def\g{\gamma}
\def\vta{\vartheta}
\def\G{\Gamma}
\def\be{\begin{equation}}
\def\ee{\end{equation}}

\def\ra{{\cal R}{\it ica}}

\def\sqr#1#2{{\vcenter{\hrule height.#2pt\hbox{\vrule width.#2pt
height#1pt \kern#1pt \vrule width.#2pt}\hrule height.#2pt}}}
\def\square{\mathchoice\sqr64\sqr64\sqr{4.2}3\sqr{3.0}3}

\def\negenspace{\kern-1.1em}\def\quer{\negenspace\nearrow}

\begin{document}

\title{Einstein-aether theory, violation of Lorentz invariance, and
  metric-affine gravity}

\author{Christian Heinicke$^{1}$\email{chh@thp.uni-koeln.de}, Peter
  Baekler$^2$\email{peter.baekler@fh-duesseldorf.de}, Friedrich W.\
  Hehl$^{1,3}$\email{hehl@thp.uni-koeln.de}}

\affiliation{$^1$Institut f\"ur Theoretische Physik, Universit\"at zu
  K\"oln, 50923 K\"oln, Germany \\$^2$Fachbereich Medien,
  Fachhochschule D\"usseldorf, University of Applied Sciences,
  Josef-Gockeln-Str.\ 9, 40474 D\"usseldorf, Germany\\ $^3$Department
  of Physics and Astronomy, University of Missouri-Columbia, Columbia,
  MO 65211, USA}

\date{2005-04-01, {\em aether19.tex}}

\begin{abstract}
  We show that the Einstein-aether theory of Jacobson and Mattingly
  (J\&M) can be understood in the framework of the metric-affine
  (gauge theory of) gravity (MAG). We achieve this by relating the
  aether vector field of J\&M to certain post-Riemannian {\it
    nonmetricity\/} pieces contained in an independent linear
  connection of spacetime. Then, for the aether, a corresponding
  geometrical curvature-square Lagrangian with a massive piece can be
  formulated straightforwardly.  We find an exact spherically symmetric
  solution of our model.
\end{abstract}

\pacs{03.50.Kk; 04.20.Jb; 04.50.+h}

\keywords{Violation of Lorentz invariance, nonmetricity, gravity,
  connection, aether, metric-affine gravity}

\maketitle

%%%%%%%%%%%%%%%%%%%%%%%%%%%%%%%%%%%%%%%%%%%%%%%%%%%%%%%%%%%%%%%%%%%%
\section{Introduction}
%%%%%%%%%%%%%%%%%%%%%%%%%%%%%%%%%%%%%%%%%%%%%%%%%%%%%%%%%%%%%%%%%%%%

In an attempt to violate Lorentz invariance locally, J\&M
\cite{JacobsonMattingly04} introduced an extra timelike 4-vector field
$u$ of unit length, see also
\cite{jacobson00,Barbero,Eling04,ElingDeser,Carroll+Lim,Graesser,Foster2005,Lim2004}
and the earlier work of Gasperini \cite{Gasperini0,Gasperini} and
Kostelecky and Samuel \cite{Kostelecky+Samuel}.  This ``aether field''
is governed by a Lagrangian carrying a kinetic term $\sim(\nabla u)^2$
and a massive term $\sim u^2$. The aether field equation is of
(massive) Yang-Mills type
\begin{equation}\nabla H+{\rm lower\; order
    \; term}\; \sim \ell\,u\,,
\end{equation}
with $H\sim \nabla u$ as field momentum (excitation) and $\ell$ as
some constant. The aether field $u$ can be considered to be an
analogue of a Yang-Mills potential $B$.

The purpose of our note is to point out that such an aether theory can
be reconstructed quite naturally as a specific example within the
framework of MAG \cite{Erice95,PRs}. The reasons are as follows:
\begin{enumerate}
\item Since the aim of the aether theory of J\&M is to violate local
  Lorentz invariance, one should abandon the (flat) Minkowski as well
  as the (curved) Riemannian spacetime, which are rigidly and locally
  Lorentz invariant, respectively, and look for a {\it post-\/}Riemannian
  structure of spacetime appropriate for such an approach. If one
  introduces in spacetime, besides the metric $g_{\a\b}$, an
  independent linear connection 1-form $\Gamma_\a{}^\b$,
  then it is known {}from literature, see \cite{PRs}, that the {\it
    nonmetricity\/}
\begin{equation}\label{structure0}
  Q_{\alpha \beta}:=-\stackrel{\Gamma}{D}\!g_{\alpha \beta }\,,
\end{equation}
where $\stackrel{\Gamma}{D}$ is the covariant exterior derivative with
respect to the connection $\Gamma_\a{}^\b$, {\it is a measure for the
  violation of local Lorentz invariance.\/} Note that the introduction
of a torsion 2-form $T^\a$ is voluntary in this context.

\item In four dimensions the nonmetricity with its 40 components can
  be split irreducibly into four pieces (see \cite{PRs}):
\begin{eqnarray}\label{Qirred}
  Q_{\a\b}&=&\,^{(1)}\!Q_{\a\b}\oplus\,^{(2)}\!Q_{\a\b}
  \oplus\,^{(3)}\!Q_{\a\b} \oplus\,^{(4)}\!Q_{\a\b}\,,\\40\hspace{6pt}
  &=& {\rm \quad 16\quad}\oplus{\rm \; 16\quad}\oplus{\rm \; \quad
    4\quad}\oplus{\rm \; \quad 4\quad}\nonumber\,.
\end{eqnarray}
In this article we want to concentrate on reconstructing the J\&M
theory in the framework of MAG. Therefore we pick the {\it two
  vector-like quantities\/} $^{(3)}\!Q_{\a\b}$ and $^{(4)}\!Q_{\a\b}$,
with four independent components each, as relevant for our purpose.
The explicit form of these two quantities will be studied below. A
result will be that the geometrical interpretation of
$^{(4)}\!Q_{\a\b}$ identifies it as equivalent to the {\it Weyl
  covector} \cite{Weyl1918} of 1918, which is related to scale
transformations (and thus extends the Lorentz to the conformal group),
whereas $^{(3)}\!Q_{\a\b}$ is related to {\it shear\/} transformations
and, accordingly, is a generic field obstructing local Lorentz
invariance.  In other words, we identify $^{(3)}\!Q_{\a\b}$ as analogue
of the aether field $u$ of J\&M.  Accordingly, by assuming an
independent linear connection of spacetime, we arrive
straightforwardly at a vector-like quantity of shear type that
dissolves local Lorentz invariance. It is part of the geometric
structure of spacetime and represents as such some kind of genuine
aether.

\item Identifying ${}^{(3)}Q_{\alpha\beta}$ as an analogue of the aether
  field $u$ suggests a Yang-Mills type Lagrangian
\begin{equation}\label{lagr}
  V_{\rm ^{(3)}Q} \sim\; \stackrel{\Gamma}{D}\!^{(3)}\!Q_{\a\b}\wedge
  {}^\star\!\!  \stackrel{\Gamma}{D}\!^{(3)}\!Q^{\a\b} +\ell\,^{(3)}
  Q_{\a\b}\wedge{}^{\star\,(3)}\!Q^{\a\b}\,,
\end{equation}
where $^\star$ denotes the Hodge star operator. The first term
parallels the kinetic aether term of J\&M, see
\cite{JacobsonMattingly04}, Eq.(2), the second one the massive term of
J\&M. If a gravitational Hilbert-Einstein Lagrangian is added to
(\ref{lagr}), we recover the basic structure of the J\&M Lagrangian.

However, we stress an important difference. In our approach we proceed
{}from variables of purely geometric origin: metric $g_{\alpha\beta}$
(defining angles and lengths), coframe $\vta^\alpha$ (representing the
local reference frame) and connection $\Gamma_\alpha{}^\beta$
(defining parallel displacement). In contrast, the aether field of
J\&M is an {\it external\/} quantity with no obvious relation to the
geometry of spacetime, it is more like a cosmological fluid. Moreover,
in spite of the presumed violation of Lorentz invariance, the J\&M
aether field $u$ has a {\it fixed\/} (timelike) {\it magnitude,} which
doesn't seem convincing to us.

In actual fact, the geometrical Lagrangian that we are going to study
below will be a bit more complicated than (\ref{lagr}). This conforms
better to the MAG models that we developed earlier. In this context we
will also demonstrate that the Lagrangian (\ref{lagr}), in spite of
the existence of the derivative terms in (\ref{lagr}), fits very well
into the first order version of MAG.

\item In the J\&M model, the coupling of the aether field to matter
  seems to be an unsolved problem. In our approach, in the sense of
  Einstein's equivalence principle, we assume {\it minimal coupling,}
  that is, partial (or exterior) derivatives are substituted by
  covariant ones: $d\longrightarrow \stackrel{\Gamma}{D}$. Thereby a
  universal coupling of matter to our aether is guaranteed.
\end{enumerate}

In Sec.II we give a sketch of metric-affine geometry, and in Sec.III
we describe the main geometrical properties of the nonmetricity.  In
Sec.IV we turn to the the different curvature 2-forms, in particular
to those that relate to post-Riemannian structures. Subsequently, in
Sec.V, we display the general form of the field equations of MAG and
compare them with those of J\&M. In Sec.VI we formulate the general
quadratic Lagrangian of MAG. We tailor it such that it becomes
somewhat analogous to the J\&M Lagrangian. We compute the
corresponding excitations (field momenta) explicitly by suitable
partial differentiation of the Lagrangian. Accordingly, the field
equations are now known explicitly.

In Sec.VII we present an exact spherically symmetric solution, the
mass and the angular momentum of which are determined in
Sec.\ref{Killing}. The prolongation techniques that we took for
finding exact solutions are explained shortly in
Sec.\ref{prolongation}. We discuss our results in
Sec.\ref{discussion}.

Our notation is as follows (see \cite{PRs,Birkbook}): We use the
formalism of exterior differential forms. We denote the frame by
$e_\a$, with the anholonomic or frame indices $\a,\b,\dots=0,1,2,3$.
Decomposed with respect to a natural frame $\partial_i$, we have
$e_\a=e^i{}_\a\,\partial_i$, where $i,j,\dots=0,1,2,3$ are holonomic
or coordinate indices. The frame $e_\a$ is the vector basis of the
tangent space at each point of the 4D spacetime manifold. The symbol
$\rfloor$ denotes the interior and $\wedge$ the exterior product. The
coframe $\vta^\b=e_j{}^\b dx^j$ is dual to the frame, i.e.,
$e_\a\rfloor \vta^\b=\delta^\b_\a$. If $^\star$ denotes the Hodge star
operator and if $\vta^{\a\b}:=\vta^\a\wedge\vta^\b$, etc., then we can
introduce the eta-basis $\eta:=\,^\star 1$,
$\>\eta^\a:=\,^\star\vta^\a$, $\>\eta^{\a\b}:=\,^\star\vta^{\a\b}$,
etc., see also \cite{Thirring}. Parentheses surrounding indices
$(\a\b) :=(\a\b+\b\a)/2$ denote symmetrization and brackets $[\a\b]
:=(\a\b-\b\a)/2$ antisymmetrization.

%%%%%%%%%%%%%%%%%%%%%%%%%%%%%%%%%%%%%%%%%%%%%%%%%%%%%%%%%%%%%%%%%%%%
\section{Metric-affine geometry of spacetime}
%%%%%%%%%%%%%%%%%%%%%%%%%%%%%%%%%%%%%%%%%%%%%%%%%%%%%%%%%%%%%%%%%%%%

Spacetime is a 4-dimensional differentiable manifold that is equipped
with a metric and a linear (also known as affine) connection (see
\cite{Schrodinger54}). The metric is of Lorentzian signature
$(-++\,+)$ and is given by
\begin{equation}\label{metric'}
  g=g_{\a\b}\,\vta^\a\otimes \vta^\b\,.
\end{equation}
In general, the coframe is left arbitrary, sometimes it is convenient
to choose it orthonormal: $g=o_{\a\b}\,\vta^\a\otimes \vta^\b$, with
$o_{\a\b}={\rm diag}(-1,+1,+1,+1)$. The linear connection $\G_\a{}^\b$
governs parallel transfer and allows to define a covariant exterior
derivative $D$ (we drop now the $\Gamma$ on top of $D$ for
convenience). We can decompose $\G_\a{}^\b$ with respect to a natural
frame:
\begin{equation}\label{connection}
  \G_\a{}^\b=\G_{i\a}{}^\b\,dx^i\,.
\end{equation} It has apparently 64
independent components. The notion of a non-trivial linear connection
is decisive in going beyond the (flat) Minkowskian spacetime. To quote
Einstein \cite{Einstein55}: ``\dots the essential achievement of
general relativity, namely to overcome `rigid' space (i.e., the
inertial frame), is {\it only indirectly} connected with the
introduction of a Riemannian metric.  The directly relevant conceptual
element is the `displacement field' ($\Gamma^l_{ik}$), which
$\,$expresses the$\,$ infinitesimal displacement of vectors. It is
this which replaces the parallelism of spatially arbitrarily separated
vectors fixed by the inertial frame (i.e., the equality of
corresponding components) by an infinitesimal operation. This makes it
possible to construct tensors by differentiation and hence to dispense
with the introduction of `rigid' space (the inertial frame). In the
face of this, it seems to be of secondary importance in some sense
that some particular $\Gamma$ field can be deduced {}from a Riemannian
metric \dots ''

The metric $g_{\a\b}$ induces a Riemannian (or Levi-Civita) connection
1-form $\widetilde{\Gamma}_{\beta}{}^{\alpha}$. In holonomic
coordinates, it reads
\begin{equation}
  \widetilde{\Gamma}_i{}^j\,
  =\,\widetilde{\Gamma}_{ki}{}^j\,dx^k,\qquad
  \widetilde{\Gamma}_{ki}{}^j\,:=\frac{1}{2}\,g^{jl}\left(\partial_ig_{kl}
    + \partial_kg_{il}-\partial_lg_{ik}\right)\,.\label{Chris}
\end{equation}
The post-Riemannian part of the connection, the {\it distortion\/}
\begin{equation}\label{distortion}
  N_\a{}^\b:=\Gamma_\a{}^\b-\widetilde{\Gamma}_\a{}^\b\,,
\end{equation}
is a tensor-valued 1-form. Its 64 independent components describes the
deviation {}from Riemannian geometry. In Einstein's theory of gravity,
general relativity (GR), spacetime is Riemannian, that is, the
distortion $N_{\a\b}$ vanishes.

\begin{table}[h]
  \caption{Gauge fields and Bianchi identities in metric-affine gravity
(MAG)}
\label{arena}
\bigskip
\begin{center}
\small
\begin{tabular}{||ll|ll|ll||}\hline\hline
\multicolumn{2}{||c|}{Potential} &
\multicolumn{2}{|c|}{Field strength} &
\multicolumn{2}{|c||}{Bianchi identity}\\
\hline
 metric& $g_{\a\b}$ & nonmetricity & $Q_{\a\b}=-Dg_{\a\b}$ & zeroth &
$DQ_{\a\b}=2R_{(\a}{}^\mu
\,g_{\b)\mu}$ \\
 coframe& $\vta^\a$ & torsion &$\hspace{6pt}T^\a=D\vta^\a$ & first
&$\hspace{6pt}DT^\a=R_\mu{}^\a\wedge \vta^\mu$ \\
 connection& $\G_\a{}^\b\quad $ & curvature &
$R_\a{}^\b=d\G_\a{}^\b-\G_\a{}^\mu\wedge\G_\mu{}^\b\quad$
& second & $DR_\a{}^\b=0$ \\
\hline\hline
\end{tabular}
\end{center}
\end{table}

There are two other measures for the deviation of a connection {}from
its Levi-Civita part: The nonmetricity $Q_{\a\b}$ of
(\ref{structure0}) and the torsion
\begin{equation}\label{torsion}
  T^\a:=D\vta^\a\,,
\end{equation}
see Table I. If one develops the covariant exterior derivative on the
right-hand-side of (\ref{structure0}) and substitutes the torsion of
(\ref{torsion}) suitably, then, after some algebra, one finds for the
distortion explicitly
\begin{equation}\label{N}
  N_{\alpha\beta}=-e_{[\alpha}\rfloor T_{\beta]} + {1\over
    2}(e_{\alpha}\rfloor e_{\beta}\rfloor
  T_{\gamma})\,\vartheta^{\gamma} + (e_{[\alpha}\rfloor
  Q_{\beta]\gamma})\,\vartheta^{\gamma} +{1\over
    2}Q_{\alpha\beta}\,.
\end{equation}
Nonmetricity and torsion can be recovered {}from $N_{\a\b}$
straightforwardly:
\begin{equation}\label{distortion2}
  Q_{\alpha\beta}=2\,N_{(\alpha\beta)}\,,\qquad T^\alpha=
  N_\beta{}^\alpha \wedge \vartheta^\beta\,.
\end{equation}
We call the (negative of the) torsion dependent piece of $N_{\a\b}$
the {\it contortion\/} 1-form $K_{\a\b}$. Like the torsion $T^\a$, it
has 24 independent components, and we have $T^\a=-K_\b{}^\a
\wedge\vta^\b$.

The torsion has 3 irreducible pieces (see \cite{PRs}), its totally
antisymmetric piece (computer name {\tt axitor}, 4 components),
\begin{equation}\label{axitor}
  ^{(3)}T^\a:=-\frac 13\,e_\a\rfloor\left(\vta^\b\wedge T_\b\right) \,,
\end{equation}
its trace ({\tt trator}, 4 components)
\begin{equation}\label{trator}
  ^{(2)}T^\a:=\frac 13\,\vta^\a\wedge T\quad{\rm with}\quad
  T:=e_\a\rfloor T^\a\,,
\end{equation}
and its tensor piece ({\tt tentor}, 16 components)
\begin{equation}\label{tentor}
  ^{(1)}T^\a:= T^\a-\, ^{(2)}T^\a -\,  ^{(3)}T^\a\,.
\end{equation}

For the Riemannian spacetime of GR, we have $T^\a=0$ and $Q_{\a\b}=0$.
If we relax the former constraint, $T^\a\ne 0$, we arrive at the
Riemann-Cartan (RC) spacetime of the viable Einstein-Cartan theory of
gravity, see \cite{eggbook,Milutin,Erice95,TrautmanEnc}. Since still
$Q_{\a\b}=0$, such a RC-spacetime carries a metric-compatible
connection and, accordingly, a length is invariant under parallel
displacement. By the same token, a RC-spacetime is locally Lorentz
invariant.  

Haugan and L\"ammerzahl \cite{HauganClausL}, see also \cite{ClausL},
argue that the presence of a torsion of spacetime violates local
Lorentz invariance. Similarly, Kostelecky \cite{Kostelecky2003} and
Bluhm and Kostelecky \cite{Bluhm+Kost} attempt to violate Lorentz
invariance already on the level of a RC-spacetime.  According to our
point of view, it is more natural to have a {\it non\/}vanishing
nonmetricity under such circumstances.  Abandoning Lorentz invariance
suggests the presence of nonmetricity, i.e., $Q_{\a\b}\ne 0$. This is
what we assume for the rest of our article.

A broad overview over the subject of violating Lorentz invariance has
been given by Bluhm \cite{Bluhm2004}, see also the references given
there.  In experiment and in observation
\cite{ClausLAdP,Mattingly2005,Muller2004,Tobar2004} there is presently
no evidence for Lorentz violations.  Nevertheless, from a theoretical
point of view (string theory, quantum gravity) a violation of Lorentz
invariance is expected at some level, for certain models, see
\cite{Allen2004,Andrianov1998}, e.g..

%%%%%%%%%%%%%%%%%%%%%%%%%%%%%%%%%%%%%%%%%%%%%%%%%%%%%%%%%%%%%%%%%%%%
\section{Some properties of the nonmetricity}
%%%%%%%%%%%%%%%%%%%%%%%%%%%%%%%%%%%%%%%%%%%%%%%%%%%%%%%%%%%%%%%%%%%%

In the presence of nonmetricity $Q_{\alpha\beta} = -Dg_{\alpha\beta}
\ne 0$, let us parallelly transport two vectors $u$ and $v$ {}from a
point $P$ along a curve with tangent vector $c$ to a neighboring point
$Q$.  The scalar product of the two vectors $g(u,v)$ will change
according to the the Lie derivative ${\cal L}_c\,g(u,v)$. With the
gauge covariant Lie derivative of a form $\psi$ (see
\cite{NesterDr,Thirring,PRs})
%\begin{equation}
${\L}_c\,\psi:=c\rfloor D\psi+D(c\rfloor \psi)\,,$
%\end{equation}
we have
\begin{eqnarray}
  {\cal L}_c\,g(u,v)&=&{\L}_c\, g(u,v)=({\L}_c\,g_{\a\b})u^\a
  v^\b=c\rfloor\left( Dg_{\a\b}\right)u^\a v^\b\,,
\end{eqnarray}
since ${\L}_c\,u^\a=0$ and ${\L}_c\,v^\a=0$ because of parallel
transfer. Thus,
\begin{equation}
  {\cal L}_c\,g(u,v)= -\left(c\rfloor Q_{\a\b}\right)u^\a v^\b= - c
  \rfloor \left[\not\!Q_{\alpha\beta} u^\alpha \, v^\beta + Q \,
    g(u,v)\right] \,,
\end{equation}
where
\begin{equation}\label{scale}
  Q:=\frac{1}{4} \, Q_\alpha{}^\alpha\,, \qquad \not\!Q_{\alpha\beta}
  := Q_{\alpha\beta} - Q \, g_{\alpha\beta} \,
\end{equation}
are the trace (Weyl covector) and the traceless (shear) part of the
nonmetricity. In the case of vanishing shear, the scalar product just
changes by a factor and the light-cone is left intact under parallel
transfer. Otherwise, with shear $\not\!Q_{\alpha\beta}$, the angle
between the vectors $u$ and $v$ does change. Hence, by admitting
nonmetricity, we dissolve the local Lorentz invariance of a
Riemann-Cartan spacetime. We depicted in Fig.1 how absolute
parallelism, length, and angles are successively abandoned.

\begin{figure}\label{fig1}
\caption{Two vectors at a point $P$ span a triangle. If
we parallelly transfer both vectors around a closed loop back to $P$,
then in the course of the round trip the triangle gets linearly
transformed.\bigskip}
\epsfig{file=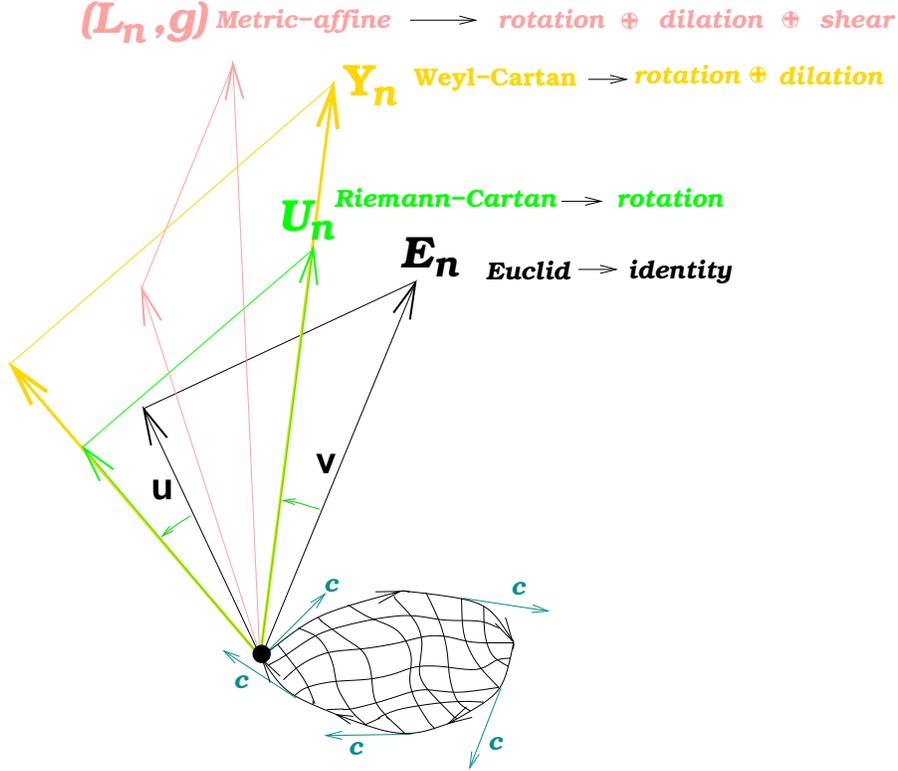,width=12cm}
\end{figure}

As we displayed in (\ref{Qirred}), the nonmetricity can be decomposed
into four pieces. We have to recapitulate some of these features. In
four dimensions, as symmetric tensor-valued $1$-form, the nonmetricity
has 40 independent components. Two vector-like pieces can be easily
identified. Firstly, the Weyl covector (\ref{scale})$_1$ can be
extracted by taking the trace of $Q_{\a\b}$. The remaining tracefree
part of the nonmetricity $\not\!\!Q_{\alpha\beta}$ in
(\ref{scale})$_2$ contains a second vector-like piece represented by
the $1$-form $\Lambda$:
\begin{equation}\label{Lambda}
\Lambda := \left(e^\beta \rfloor\!
\qslash_{\alpha\beta}\right) \, \wedge
\vartheta^\alpha \,.
\end{equation}
The 2-form\footnote{For $n$ dimensions, we have $ P _{\alpha} :=
  \qslash_{\alpha\beta} \wedge \vartheta^\beta - \frac{1}{n-1} \,
  \vartheta_\alpha \wedge \Lambda$. In \cite{PRs} we introduced a
  2-form $\Omega_\a$ instead that is related to $P_\a$ as follows:
  $P_\a=(-1)^{{\rm Ind}(g)}\,^\star\Omega_\a$.}
\begin{equation}\label{Omega}
 P _{\alpha} := \qslash_{\alpha\beta} \wedge
  \vartheta^\beta - \frac{1}{3} \, \vartheta_\alpha \wedge \Lambda\,,
\end{equation}
with the properties
\begin{equation}
 P^{\alpha}\wedge {\vartheta}_{\alpha}  =  0\,,\qquad
e_{\alpha}\rfloor P^{\alpha}  = 0\,,
\end{equation}
turns out to be related to a further irreducible piece of $Q_{\a\b}$.
We have (see \cite{PRs})
\begin{eqnarray}\label{deco4}
{}^{(4)}Q_{\alpha\beta} & := & Q \, g_{\alpha\beta} \,,\quad\\
  {}^{(3)}Q_{\alpha\beta} & := & \frac{4}{9} \left(
    \vartheta_{(\alpha} \,e_{\beta )}\rfloor - \frac{1}{4} \,
    g_{\alpha\beta} \right)\Lambda \,,\label{3Q} \\
  {}^{(2)}Q_{\alpha\beta} & := & -\frac{2}{3} \, e_{(\alpha} \rfloor
   P _{\beta)}\label{Q2} \,, \\
{}^{(1)}Q_{\alpha\beta} &:=& Q_{\alpha\beta} -
  {}^{(2)}Q_{\alpha\beta} - {}^{(3)}Q_{\alpha\beta} -
  {}^{(4)}Q_{\alpha\beta} \,.\label{deco1}
\end{eqnarray}
As we want to relate the aether vector field $u$ of J\&M to
the nonmetricity, it is obvious that $^{(4)}Q_{\alpha\beta}\sim Q$ and
$^{(3)}Q_{\alpha\beta}\sim \Lambda$ are the objects of our main
interest.

The nonmetricity $Q_{\a\b}=Q_{i\a\b}\,dx^i$, {}from a geometrical point
of view, can be understood as a {\it strain\/} measure for the
different directions specified by the 1-forms $dx^i$. In accordance
with what we stated above, (\ref{scale})$_2$ defines a shear measure
$\not\!\!Q_{\alpha\beta}$ since the dilation measure $Q$ is subtracted
out. It is then immediately clear that $^{(4)}Q_{\alpha\beta}\sim Q$
is related to {\it dilations\/} and $^{(3)}Q_{\alpha\beta}\sim
\Lambda$ to {\it shears\/}. Therefore, generically it is the 1-form
$\Lambda$ that is related to the aether vector $u$.  However, we will
keep also the Weyl 1-form $Q$ since tentatively it seems to be
related to the constraint $u^2=1$ of J\&M.

Later, for the prolongation of previously known solutions, we need the
expression $Q_{\alpha\beta}\wedge\vartheta^\beta$. It is useful to
express it also in terms of $Q$, $\Lambda$, and $P_\a$. If we
substitute the irreducible decomposition (\ref{Qirred}) into
$Q_{\alpha\beta}\wedge\vartheta^\beta$ and remember (\ref{deco4}) to
(\ref{deco1}), we find
\begin{equation}\label{qwetchtheta}
  Q_{\alpha\beta}\wedge\vartheta^\beta = P_\alpha - \frac{1}{3} \,
 ( \Lambda -3 Q)\wedge \vartheta_\alpha \,.
\end{equation}
Note that $^{(1)}Q_{\alpha\beta}\wedge {\vartheta}^{\beta}=0$, whereas
the other irreducible pieces contribute.

As we saw in (\ref{lagr}), we need ``massive'' terms in the
nonmetricity for the construction of a J\&M type Lagrangian. Since
${}^\star {}^{(4)}Q^{\alpha\beta}= g_{\a\b}\, ^\star Q$, we find
straightforwardly
\begin{equation}\label{q4square}
 ^{(4)}Q_{\alpha\beta} \wedge{}^{\star\,(4)}Q^{\alpha\beta}
= 4Q\wedge {} ^\star Q\,.
\end{equation}
In the case of $^{(3)}Q_{\alpha\beta}$, things are a bit more
complicated. However, if we use the formulas
$\vta^\a\wedge\,^\star\vta_\b=\delta^\a_\b\,\eta$ and $\eta=\,^\star 1$,
then, after some algebra, we arrive at the same type of formula
\begin{equation}\label{q3square}
  {}^{(3)}Q_{\alpha\beta} \wedge \, {}^{\star\,(3)}Q^{\alpha\beta} =
  \frac{4}{9} \Lambda \wedge{} ^\star \Lambda \,.
\end{equation}
We find the simplicity of (\ref{q4square}) and (\ref{q3square})
remarkable.

%%%%%%%%%%%%%%%%%%%%%%%%%%%%%%%%%%%%%%%%%%%%%%%%%%%%%%%%%%%%%%%%%%%%
\section{Curvature}
%%%%%%%%%%%%%%%%%%%%%%%%%%%%%%%%%%%%%%%%%%%%%%%%%%%%%%%%%%%%%%%%%%%%

The standard definition of the curvature 2-form $R_\a{}^\b$ is given
in Table I. The Riemannian curvature (of GR) we denote by
$\widetilde{R}_\a{}^\b$; of course, $\widetilde{R}_{\a\b}=-
\widetilde{R}_{\b\a}$. The curvature $R_\a{}^\b$ can be decomposed
into its symmetric and antisymmetric part according to
\begin{equation}
  R_{\a\b}=W_{\a\b}+Z_{\a\b}\,,\quad{\rm with}\quad
  W_{\a\b}:=R_{[\a\b]}\,,\; Z_{\a\b}:=R_{(\a\b)}\,.
\end{equation}
The ``rotational'' curvature $W_{\a\b}$ characterizes a RC-space,
whereas the ``strain'' curvature $Z_{\a\b}$ only emerges if
nonmetricity is admitted. Therefore, the investigation of the ``Lorentz
violating'' curvature $Z_{\a\b}$ is of central importance to our
paper.

However, in order to link up our theory to GR and to the
Einstein-Cartan theory with its RC-spacetime, we have to take a look
at the rotational curvature $W_{\a\b}$ as well. It can be decomposed
into six irreducible pieces:
\begin{eqnarray}\label{Wdeco}
  W_{\a\b}&=&\,^{(1)} W_{\a\b} \oplus\,^{(2)} W_{\a\b} \oplus\,^{(3)}
  W_{\a\b} \oplus\,^{(4)} W_{\a\b} \oplus\,^{(5)} W_{\a\b}
  \oplus\,^{(6)} W_{\a\b}\\ &=&\hspace{10pt}{\tt weyl}\oplus{\tt
    paircom}\oplus{\tt pscalar}\oplus{\tt ricsymf}\oplus{\tt ricanti}
  \oplus{\tt scalar}\,,\nonumber\\ 36&=&\hspace{14pt}
  10\hspace{14pt}\oplus \hspace{14pt}
  9\hspace{14pt}\oplus\hspace{14pt}1
  \hspace{14pt}\oplus\hspace{14pt}9\hspace{14pt}
  \oplus\hspace{14pt}6\hspace{14pt}\oplus\hspace{14pt}1\,.\nonumber
\end{eqnarray}
The names are those that we use in our computer algebra programs
\cite{Socorro} for the decomposition of $W_{\a\b}$. In a RC-space, all
six pieces in (\ref{Wdeco}) are nonvanishing in general. If torsion
$T^\a=0$, then, as can be seen {}from the first Bianchi identity in
Table I, {\tt paircom}, {\tt pscalar}, and {\tt ricanti} vanish and we
are left with the three pieces known {}from GR: The Weyl tensor {\tt
  weyl}, the symmetric tracefree Ricci tensor {\tt ricsymf}, and the
curvature scalar {\tt scalar}. Thereby the curvature reduces {}from the
36 independent components in a RC-space to 20 independent components
in a Riemannian space, a result well-known {}from GR.

Let us come back to the strain curvature $Z_{\a\b}=Z_{\b\a}$.
Obviously, it has one distinctive piece, namely its trace
$Z:=g^{\a\b}Z_{\a\b}=Z_\g{}^\g$. It should be noted that $Z$ is
related to a {\it premetric\/} quantity. In a space in which only a
linear connection is specified, the curvature $R_\a{}^\b$ can be
contracted, $R_\g{}^\g$, even if a metric is not present. Thus
$R_\g{}^\g$ and, as a consequence, also $Z$ is rightfully called {\tt
  dilcurv}, the part of the curvature related to dil(at)ations. This
is an irreducible piece of $Z_{\a\b}$ and we call it
\begin{equation}\label{Z4}
^{(4)}Z_{\a\b}:=\frac{1}{4}\,g_{\a\b}Z\,.
\end{equation}
Since on the level of the nonmetricity dilations are related to
${}^{(4)}Q_{\alpha\beta}$, we denoted the related curvature piece by
the same number. In fact, the zeroth Bianchi identity in Table I, if
contracted, yields $g^{\a\b}DQ_{\a\b}=2Z_\g{}^\g=Z$. By partial
integration, we find
\begin{equation}\label{Z4Q4}
  Z=2dQ\qquad{\rm or}\qquad ^{(4)}Z_{\a\b}= \frac{1}{2} \,
  g_{\alpha\beta} \, dQ=\frac 12\left(D\,^{(4)}Q_{\alpha\beta}
    +\qslash_{\a\b}\wedge Q \right)\,.
\end{equation}
Apparently, to the field strength $Z$, that is, to {\tt dilcurv},
there belongs the potential $Q$, the Weyl covector.\footnote{Let us
  recall, it was {\tt dilcurv} that Weyl \cite{Weyl1918} used in his
  unsuccessful unified field theory of 1918 to describe the
  electromagnetic field strength $F$ --- and $Q$ was meant to be the
  electromagnetic potential $A$, see Goenner \cite{Goenner} and
  O'Raifeartaigh \cite{Lochlainn}.}

The tracefree part of the strain curvature
\begin{equation}\label{Zslash}
  \not\!Z_{\a\b}:=Z_{\a\b}-\frac 14\,Zg_{\a\b}\,
\end{equation}
represents the shear curvature. In terms of $\not\!Z_{\a\b}$, a
further decomposition of $Z_{\a\b}$ is possible. We have collected the
results in Appendix \ref{straincurv}. We find a decomposition of
$Z_{\a\b}$ into five irreducible pieces:
\begin{eqnarray}
  Z_{\a\b}&=&\,^{(1)}\! Z_{\a\b}\oplus \,^{(2)}\!
  Z_{\a\b}\oplus\,^{(3)}\!  Z_{\a\b}\oplus\,^{(4)}\!
  Z_{\a\b}\oplus\,^{(5)}\! Z_{\a\b}\,,\\ 60\hspace{5pt}&
  =&\hspace{11pt}30\hspace{11pt} \oplus\hspace{11pt} 9\hspace{11pt}
  \oplus\hspace{11pt}6\hspace{11pt} \oplus\hspace{11pt}
  6\hspace{11pt}\oplus\hspace{11pt}9\,.\nonumber
\end{eqnarray}

Alerted by our results with respect to the nonmetricity, we expect
that $^{(3)}\!Z_{\a\b}$ with its six independent components may be
related to $^{(3)}Q_{\a\b}$ in a similar way as $^{(4)}\!Z_{\a\b}$ is
to $^{(4)}Q_{\a\b}$, see (\ref{Z4Q4}). Symbolically, we expect
\begin{eqnarray}\label{corresp4}
  \text{Weyl 1-form}\quad Q\sim
  \,^{(4)}Q\,&\stackrel{d}{\longrightarrow}& \,^{(4)}Z\sim dQ\,,\\
  \label{corresp3} \text{shear 1-form}\quad\Lambda\sim
  \,^{(3)}Q\,&\stackrel{d}{\longrightarrow}& \,^{(3)}Z\sim d\Lambda\,.
\end{eqnarray}
The rigorous form of relation (\ref{corresp4}) was already presented
in (\ref{Z4Q4}).  What about (\ref{corresp3})? Well, life is a bit
more complicated than (\ref{corresp3}) suggests. If we take the
definition of $^{(3)}Z_{\a\b}$ {}from (\ref{Z3de}) and the zeroth
Bianchi identity $DQ_{\a\b}=2Z_{\a\b}$ {}from Table I, then, after
some light algebra, we find
\begin{equation}\label{Z3'}
  ^{(3)}\!Z_{\a\b}= {1\over 3}\,\left( 2\,\vta_{(\a}\wedge
    e_{\b)}\rfloor-g_{\a\b}\right)\hat{\Delta}\,,
\end{equation}
with
\begin{equation}\label{Z3''}
  \hat{\Delta}=\frac 14\,\vta^\a\wedge e_\b\rfloor\left(
    DQ_{\a\b}-g_{\a\b}DQ\right)=\frac 14\,\vta^\a\wedge
  e_\b\rfloor\left( D\not\!Q_{\a\b}+Q\wedge\not\!Q_{\a\b}\right)\,.
\end{equation}
Apparently, Eq.(\ref{Z3''}) turns out to be more complicated
than we guessed in (\ref{corresp3}). We can only hope to find a
formula of type (\ref{corresp3}), if we forbid certain pieces of the
connection to occur.

In Appendix \ref{hatDelta} we will show that under the conditions
\begin{equation}\label{circ}
^{(2)}\!Q_{\a\b}=0\,,\qquad ^{(1)}\!T^\a=\,^{(3)}\!T^\a=0\,,
\end{equation}
we have, see (\ref{B15}) and (\ref{Z3'}),
\begin{equation}\label{3Ztrunc}
  \hat{\Delta}\stackrel{(\ref{circ})}{=}\frac 16\,d\Lambda\quad{\rm
    or}\quad\, ^{(3)}\!Z_{\a\b}\stackrel{(\ref{circ})}{=}{1\over
    18}\,\left( 2\,\vta_{(\a}\wedge
    e_{\b)}\rfloor-g_{\a\b}\right)d\,\Lambda\,.
\end{equation}
Accordingly (\ref{corresp3}) turns out to be correct after all,
provided the conditions (\ref{circ}) are met. Thus, modulo the
conditions (\ref{circ}), our goal is reached of constructing a gauge
Lagrangian \`a la (\ref{lagr}) in terms of $\Lambda$ and $Q$. For
$^{(4)}Q_{\a\b}$ we have, see (\ref{Z4}) and (\ref{Z4Q4}),
\begin{equation}\label{ident4'}
 {}^{(4)}Z_{\alpha\beta} \wedge  {}^{\star\,(4)}Z^{\alpha\beta} =
  \frac{1}{4} Z \wedge {}^\star  Z =  dQ\wedge {}^\star dQ \,,
\end{equation}
and for $^{(3)}Q_{\a\b}$, using (\ref{C1}) and (\ref{3Ztrunc})$_1$,
\begin{equation}\label{ident3'}
  {}^{(3)}Z_{\alpha\beta}\wedge {}^{\star\,(3)}Z^{\alpha\beta} =
  \frac{4}{3} \hat{\Delta}\wedge{}^\star \hat{\Delta}
  \stackrel{(\ref{circ}) }{=}\frac{1}{27} d\Lambda\wedge
   {}^\star d\Lambda \,.
\end{equation}

%%%%%%%%%%%%%%%%%%%%%%%%%%%%%%%%%%%%%%%%%%%%%%%%%%%%%%%%%%%%%%%%%%%%
\section{Currents and field equations of MAG, comparison with J\&M}
%%%%%%%%%%%%%%%%%%%%%%%%%%%%%%%%%%%%%%%%%%%%%%%%%%%%%%%%%%%%%%%%%%%%

Similar as J\&M \cite{JacobsonMattingly04}, we assume gravity and
aether to exist, and then we couple them minimally to certain matter
fields $\Psi$. However, in our case the gauge potentials ($
g_{\alpha\beta},\, \vartheta^{\alpha},\,\Gamma_{\alpha}{}^{\beta}$)
represent ordinary gravity ($g_{\alpha\beta},\, \vartheta^{\alpha}$)
as well as the image ($\Gamma_{\alpha}{}^{\beta}$) of the J\&M aether,
and they are both part of the geometry of spacetime. The total
first-order Lagrangian reads
\begin{equation}\label{totlagr}
L_{\rm tot}= V(g_{\alpha\beta}, \vartheta^{\alpha}, Q_{\alpha\beta},
  T^{\alpha}, R_{\alpha}{}^{\beta}) + L(g_{\alpha\beta},
  \vartheta^{\alpha},\Psi , D\Psi)\,.
\end{equation}
The independent variables of the action principle are $
g_{\alpha\beta},\, \vartheta^{\alpha},\,\Gamma_{\alpha}{}^{\beta}$,
and $\Psi$. The variation of the matter Lagrangian
\begin{equation}\label{varlagr}
\delta L =
{1\over2}\delta g_{\alpha\beta}\,\sigma^{\alpha\beta} +
\delta\vartheta^{\alpha}\wedge\Sigma_\alpha+
\delta\Gamma_\alpha{}^{\beta}
\wedge\Delta^\alpha{}_\beta +\delta\Psi\wedge {{\delta L}\over{\delta
\Psi}}
\end{equation}
allows us identify the material currents coupled to the potentials as
metric and canonical energy-momentum and as hypermomentum,
respectively: ($\sigma_{\a\b},\Sigma_\a,\Delta^\a{}_\b$).  The
energy-momenta $\sigma_{\a\b}$ and $\Sigma_\a$ are related to each
other by a Belinfante-Rosenfeld type of relation. The hypermomentum
splits in spin current $\oplus$ dilation current $\oplus$ shear
current:
\begin{equation}\label{hyper}
  \Delta_{\alpha\beta}= {\tau_{\alpha\beta}}+ {1\over
    4}\,g_{\alpha\beta}\,\Delta^\g{}_\g+
  {\buildrel\frown\over{\Delta\quer}}_{\alpha\beta}\,,\qquad
  \tau_{\a\b}=-\tau_{\b\a}\,,\quad
  {\buildrel\frown\over{\Delta\quer}}_{\alpha\beta}=
  {\buildrel\frown\over{\Delta\quer}}_{\beta\alpha}\,,\;
  {\buildrel\frown\over{\Delta\quer}}{}^{\,\g}{}_\g=0\,.
\end{equation}
The hypothetical shear current ${\buildrel\frown\over{\Delta
    \quer}}_{\alpha\beta}$ is discussed in \cite{YuvalCQG,HehlObukhov97}, see
also the literature given there.

Our strategy is to leave open the explicit form of the gauge
Lagrangian $V$ for the time being and to introduce the excitations (or
field momenta) of the gauge fields instead:
\begin{equation}\label{excitations}
  M^{\alpha\beta} = - 2\,{{\partial V}\over{\partial
      Q_{\alpha\beta}}}\,,\quad H_{\alpha} = - {{\partial
      V}\over{\partial T^{\alpha}}}\,,\quad H^{\alpha}{}_{\beta} = -
  {{\partial V}\over{\partial R_{\alpha}{}^{\beta}}}\, .
\end{equation}
These three {\it constitutive\/} laws, expressing the excitations in
terms of the field strengths, characterize the physical properties of
the spacetime continuum under consideration.

The field equations read \cite{PRs}
\begin{eqnarray}\label{zeroth}
  DM^{\alpha\beta} - m^{\alpha\beta} &=& \sigma^{\alpha\beta}
  \qquad\,\qquad {\rm (zeroth)}\,,\\ DH_{\alpha} -
  E_{\alpha}\hspace{6pt} & =& \Sigma_{\alpha} \qquad\>\;\qquad {\rm
    (first)}\,,\label{first}\\ DH^{\alpha} {}_{\beta} -
  E^{\alpha}{}_{\beta} &=& \Delta^{\alpha}{}_{\beta} \qquad\qquad{\rm
    (second)}\,,\label{second}\\ {{\delta L}\over{\delta\Psi}} &=&
  0\quad\qquad\qquad\,\; {\rm (matter)}\,.\label{matter}
\end{eqnarray}
If the second field equation (\ref{second}) is fulfilled, then either
the zeroth field equation (\ref{zeroth}) or the first one
(\ref{first}) is redundant due to some Noether identities. Hence we
need only to consider (\ref{zeroth}),(\ref{second}),(\ref{matter}) or
(\ref{first}),(\ref{second}),(\ref{matter}).

On the right-hand-side of each of the gauge field equations
(\ref{zeroth}) to (\ref{second}) we have a material current, on the
left-hand-side first the Yang-Mills type term ``derivative of
excitation'' minus, as second term, a gauge current that, together
with the material current, features as source of the corresponding
gauge field. The gauge currents turn out to be \cite{PRs} the metrical
energy-momentum of the gauge fields
\begin{equation}\label{zerothx} m^{\alpha\beta} :=
  2\,{{\partial V}\over{\partial g_{\alpha\beta}}}=
  \vartheta^{(\alpha} \wedge E^{\beta)} + Q^{(\a}{}_\g\wedge
  M^{\b)\g}-T^{(\alpha}\wedge H^{\beta)} - R_\g {}^{(\alpha} \wedge
  H^{|\g|\beta)} + R^{(\a|\gamma|} \wedge H^{\b)} {}_\g \,,
\end{equation}
the canonical energy-momentum of the gauge fields\footnote{For the
  relations between different energy-momentum currents in
  gravitational theory one should also compare Itin \cite{Yakov}.}
\begin{equation}\label{firstx} E_{\alpha} :=
  {{\partial V}\over{\partial\vartheta^{\alpha}}}= e_{\alpha}\rfloor V
  + (e_{\alpha}\rfloor T^{\beta})\wedge H_{\beta} + (e_{\alpha}\rfloor
  R_{\beta}{}^{\gamma})\wedge H^{\beta}{}_{\gamma} +
  {1\over2}(e_{\alpha}\rfloor Q_{\beta\gamma})\, M^{\beta\gamma}\,,
\end{equation}
and the hypermomentum of the gauge fields
\begin{equation}\label{secondx}
  E^{\alpha}{}_{\beta}:= {{\partial
      V}\over{\partial\Gamma_{\alpha}{}^{\beta}}}= -
  \vartheta^{\alpha}\wedge H_{\beta} - g_{\beta\gamma}\,
  M^{\alpha\gamma}\,,
\end{equation}
respectively.

Like J\&M, we will concentrate on the sourcefree region, that is, we
assume that the material currents vanish. We discussed them here
in order to find the physical interpretations of the gauge currents
$m^{\a\b},\,E_\a$, and $E^\a{}_\b$. Since J\&M consider a symmetric
energy-momentum in their theory, we consider the sourcefree zeroth and
second field equations:
\begin{eqnarray}\label{zeroth'}
  DM^{\alpha\beta}& =& m^{\alpha\beta}\,,\\
  DH^{\alpha} {}_{\beta} &=& E^{\alpha}{}_{\beta}\,.\label{second'}
\end{eqnarray}
These are the two field equations that underlie our model of the J\&M
theory. The rest of the paper will be devoted to making them explicit
and for finding exact solutions of them.

The field equation (\ref{zeroth'}) is of an Einsteinian type. If the
gauge Lagrangian $V$ depends on a Hilbert-Einstein term $R_{\rm sc}$,
inter alia (curvature {\it sc\/}alar), then, within $m^{\a\b}$, the
Einstein 3-form emerges. The left hand side of (\ref{zeroth'}) should
depend on the shear 1-form $\Lambda$. This can be achieved by putting
$M^{\a\b}=-2\partial V/\partial Q_{\a\b}$ proportional to
$^{(3)}Q^{\a\b}$, see (\ref{3Q}).  As a consequence, the Lagrangian
$V$ carries a quadratic $^{(3)}Q$ piece and we expect $V\sim R_{\rm
  sc}+\, ^{(3)}Q^2$. Under these circumstances, Eq.(\ref{zeroth'}) is
the analogue of \cite{JacobsonMattingly04}, Eq.(5).  Both equations
have 10 independent components.\footnote{In our approach we don't find
  the analogue of the $\nabla( Ju)$ terms in the first line of the
  aether stress tensor of J\&M \cite{JacobsonMattingly04}, Eq.(9). The
  reason is clear. In the J\&M aether these terms arise, see
  \cite{Eling04}, p.2, ``{}from varying the metric dependence of the
  connection.''  However, in our case the connection is an independent
  variable.  Incidentally, an aether stress tensor that depends, as in
  the J\&M theory, on second derivatives of the field variable is not
  particularly plausible anyway.}

Our second field equation (\ref{second'}), with the gauge
hypermomentum as source, has 64 independent components. For this
reason, as we argued above, we have to kill all components apart {}from
those 4 components related to the shear 1-form
$\Lambda\sim\,^{(3)}Q_{\a\b}$. At a first glance, this seems to be an
impossible task. A little reflection shows that the situation is not
hopeless at all.  Substitute (\ref{secondx}) into (\ref{second'}):
\begin{equation}\label{second''}
  DH^{\alpha} {}_{\beta}= - \vartheta^{\alpha}\wedge H_{\beta} -
  M^{\alpha}{}_\b\,.
\end{equation}
According to the the last paragraph, we have $M^{\a\b}\sim
\,^{(3)}Q^{\a\b}$, that is, $ M^{\alpha}{}_\b$ on the right hand side
of (\ref{second''}) depends on 4 independent components.  The term
with $H_\a=-\partial V/\partial T^\a$ can be chosen to vanish by
forbidding explicit torsion dependent terms in the Lagrangian.

Left over for discussion is $H^{\alpha} {}_{\beta}=-\partial
V/\partial R_\a{}^\b$ on the left hand side of (\ref{second''}).
Clearly we want this term to depend in an essential way on the shear
1-form $\Lambda$. A look at (\ref{ident3'}) convinces us to take
$H^{\alpha} {}_{\beta}\sim {}^{(3)}Z^{\alpha}{}_{\beta}\sim d\Lambda$.
Then the left hand side of (\ref{second''}) becomes a wave type
expression $\sim \square\,\Lambda$ with four essential components.
Hence our equation reduces to just four components, like the
corresponding J\&M equation. Thus, our second field equation
(\ref{second'}) is undoubtedly the analogue of
\cite{JacobsonMattingly04}, Eq.(4).  At the same time it is also clear
that we could introduce a richer aether structure than the one J\&M
studied by means of their vector field $u$.

If we collect our heuristic arguments for constructing the MAG
analogue of the J\&M aether, then we arrive at
\begin{equation}\label{toy0}
  V_{\rm J\&M}\sim\frac{1}{\kappa}( R_{\rm sc}+{}
  ^{(3)}Q^2)+\frac{1}{\rho}\,^{(3)}Z^2\,,
\end{equation}
with $\kappa$ as gravitational constant and $\rho$ as a dimensionless
coupling constant. This first toy Lagrangian should be compared with
\cite{JacobsonMattingly04}, Eq.(1). Our massive term $^{(3)}Q^2$
resemble the constraint piece in the J\&M Lagrangian, whereas
\begin{equation}\label{3Z}
  ^{(3)}Z^2 \sim d\Lambda \wedge  {}^\star d\Lambda
  \sim g^{\g[\alpha}\,g^{\beta]\delta} \, (\partial_\alpha
  \Lambda_\beta)\,(\partial_\gamma \Lambda_\delta) \, \eta
\end{equation}
is the analogue of $K^{ab}{}_{mn}\nabla_au^m\nabla_bu^n$.  However,
J\&M have, in $K^{ab}{}_{mn}$, four open constants $c_1, c_2, c_3,
c_4$.  In the specific model investigated below, we concentrate on the
simple Maxwell-type kinetic term (57). In doing so, we seem closer to
Kostelecky and Samuel \cite{Kostelecky+Samuel} than to J\&M.  However,
the general MAG Lagrangian, see Appendix \ref{MAGLagrangian},
Eq.(\ref{QMA}), encompasses kinetic terms of all 11 irreducible pieces
of the curvature and thereby generalizes the 4 parameters of J\&M
considerably.

Let us look back to our first ansatz for an aether Lagrangian in
Eq.(\ref{lagr}). There we had derivative terms of the nonmetricity.
However, such terms are not allowed in first order MAG, see the gauge
Lagrangian $V$ in (\ref{totlagr}). Only an algebraic dependency of the
field strengths $Q_{\a\b},T^\a,R_\a{}^\b$, mostly quadratic for
dimensional reasons, is allowed. Nevertheless, by the zeroth Bianchi
identity, see Table I, the derivatives can be removed and transformed
to terms algebraic in the curvature; this is at least possible for
$\,^{(3)}\!Q_{\a\b}$ and $\,^{(4)}\!Q_{\a\b}$. Thus also the
Lagrangian (\ref{lagr}) falls into the category of allowed Lagrangians
within MAG. At the same time we see again how closely nonmetricity and
shear curvature are interwoven.

We would like to stress that the formalism of MAG that we developed in
this paper up to now is exact and free of any hand waving arguments.
It is a first order Lagrange-Noether gauge formalism of non-Abelian
nature and our gauge field equations (\ref{zeroth'}) and
(\ref{second'}) are coupled nonlinear partial differential equations
of second order in the gauge potentials $g_{\a\b}\,,
\,\Gamma_\a{}^\b$. The only hand waving is involved in the {\it
  explicit\/} choice of the gauge Lagrangian $V(g_{\alpha \beta},
\vartheta^{\alpha}, Q_{\alpha\beta}, T^{\alpha},
R_{\alpha}{}^{\beta})$ in (\ref{totlagr}). Since we want to develop a
model that is an image of the J\&M theory, we have to hand pick a
suitable $V$. In (\ref{toy0}) we made a first attempt.

%%%%%%%%%%%%%%%%%%%%%%%%%%%%%%%%%%%%%%%%%%%%%%%%%%%%%%%%%%%%%%%%%%%%
\section{A Lagrangian for gravity and aether}
%%%%%%%%%%%%%%%%%%%%%%%%%%%%%%%%%%%%%%%%%%%%%%%%%%%%%%%%%%%%%%%%%%%%

In gauge theories the Lagrangian is assumed to be quadratic in the
field strengths, in our case in $R_\a{}^\b\,,\,T^\a\,,$ and
$Q_{\a\b}$. The most general parity conserving Lagrangian of such
a type has been displayed in Appendix \ref{MAGLagrangian}. For
modeling the J\&M aether theory, we don't need this very complicated
expression in its full generality. Nevertheless, let us look at its
basic structure:
\begin{equation}\label{toy1}
  V_{\rm MAG}\sim\frac{1}{\kappa}( R_{\rm sc}+ \lambda_0 + T^2 + TQ +
  Q^2)+\frac{1}{\rho}(W^2+Z^2)\,.
\end{equation}
All indices are suppressed. The expression in the first parentheses
describes $(g_{\a\b},\vta^\a)$-gravity of the Newton-Einstein type,
including a cosmological term with $\lambda_0$.  This ``weak'' gravity
is governed by the conventional gravitational constant $\kappa$. If
only these terms are present, the propagation of $\Gamma_\a{}^\b$ is
inhibited.  Then, in addition to conventional gravity, only new {\it
  contact\/} interactions emerge that are glued to matter. The {\it
  viable} Einstein-Cartan theory with its spin-spin contact
interaction is an example.

If one desires to make the connection $\Gamma_\a{}^\b$ propagating,
see \cite{Peter,Kuh} for vanishing and
\cite{Babourova,Floreanini,Dereli1,TuckerWang1,TuckerWang2,Dima1,Dima2}
for nonvanishing nonmetricity, then one has to allow for
curvature-square pieces $W^2$ and $Z^2$, as shown in (\ref{toy1}).
This ``strong gravity'', the potential of which is $\Gamma_\a{}^\b$,
is governed by a new dimensionless coupling constant $\rho$. Our
hypothesis is that such a universal strong gravitational interaction
is present in nature.

The J\&M aether, in our interpretation, allows at least the
$^{(3)}Q_{\a\b}$ piece of the connection to propagate, see the
$^{(3)}Z^2$ piece in the ansatz (\ref{toy0}).  We would like to stick
to (\ref{toy0}) as closely as possible. Since we search for an exact
spherically symmetric solution of our model to be defined, we need to
be flexible in the exact choice of the Lagrangian. After some
computer algebra {\it experiments,} we came up with the following
toy Lagrangian that we are going to investigate in the context of the
J\&M aether theory:
%\begin{eqnarray}\label{toy2}
%  V &=& \frac{1}{2\kappa}\, \left[-a_{0}\, \left(
%      R^{\alpha\beta}\wedge\eta_{\alpha\beta}
%      +2{\lambda}_{0}\,{\eta}\,\right) + a_2 \, T^{\alpha} \wedge\,
%^\star\,^{(2)}T_\alpha +  2  c_{3}\,^{(3)}Q_{\alpha\beta}
%    \wedge{\vartheta}^{\alpha}\wedge{} ^{\star }\!\, T^{\beta}\right.
%  \nonumber\\& +&\left. Q_{\alpha\beta} \wedge {}^\star \left(
%    b_1 {}^{(1)}Q^{\alpha\beta}+b_3 \, {}^{(3)}Q^{\alpha\beta}  \right)
%\right] - \frac{1}{2}\,R^{\alpha \beta} \wedge{}^{\star
%  }\! \left( w_{6}\,^{(6)}W_{\alpha\beta}+
%  {z}_{3}\,^{(3)}Z_{\alpha\beta}\right)\,.
%\end{eqnarray}
\begin{eqnarray}\label{toy2}
  V &=& \frac{1}{2\kappa}\, \left[-a_{0}\, \left(
      R^{\alpha\beta}\wedge\eta_{\alpha\beta}
      +2{\lambda}_{0}\,{\eta}\,\right) + Q_{\alpha\beta} \wedge
    {}^\star \left( b_1 {}^{(1)}Q^{\alpha\beta}+b_3 \,
      {}^{(3)}Q^{\alpha\beta} \right) \right]
  \nonumber\\&&\hspace{-8pt} - \frac{ {z}_{3}}{2\rho}\,R^{\alpha \beta}
  \wedge{}^{\star }\!  \,^{(3)}Z_{\alpha\beta}\,.
\end{eqnarray}
%We put $\rho=1$. We have the following constants of order unity: $a_0,
%a_2, c_3, b_1, b_3, w_6, z_3$.
We have the following constants of order unity: $a_0, b_1, b_3, z_3$.
As compared to ({\ref{toy0}), we have no torsion piece. However, we
  added in a $^{(1)}Q^{\alpha\beta}$ piece.  The most general quadratic
  Lagrangian (\ref{QMA}) is appreciably more complicated than
  (\ref{toy2}). Nevertheless, without our computer algebra programs
  (see \cite{Socorro}, \cite{Grab}, \cite{Birkbook}) we would not have
  been able to handle the messy expressions.

Once the Lagrangian is specified, it is simple to calculate the gauge
excitations by {\it partial\/} differentiation of (\ref{toy2}) with
respect to $Q_{\a\b}$, $T^\a$, and $R_\a{}^\b$:
\begin{eqnarray}\label{Mexcit}
  M^{\alpha\beta} &=& -\frac{2}{\kappa} \,^\star \left(b_1
      \, {}^{(1)}Q^{\alpha\beta}+b_3
      \,{}^{(3)}Q^{\alpha\beta}\right) \,, \\
  \label{H1excit} H_\alpha &=&\hspace{10pt} 0\,, \\
  \label{H2excit} H^\alpha{}_\beta &=& \hspace{8pt}\frac{a_0}{2\kappa} \,
  \eta^\alpha{}_\beta + \frac{z_3}{\rho}\,^\star
  {}^{(3)}Z^\alpha{}_\beta\,.
\end{eqnarray}
These excitations have to be substituted into the field equations
(\ref{zeroth'}),(\ref{second'}) and into the gauge currents
(\ref{zerothx}),(\ref{firstx}),(\ref{secondx}), respectively.
\bigskip

%%%%%%%%%%%%%%%%%%%%%%%%%%%%%%%%%%%%%%%%%%%%%%%%%%%%%%%%
\section{A simple spherically symmetric solution of MAG}
\label{simplesolution}
%%%%%%%%%%%%%%%%%%%%%%%%%%%%%%%%%%%%%%%%%%%%%%%%%%%%%%%%

We look for exact spherically symmetric solutions of the field
equations belonging to the Lagrangian (\ref{toy2}). For this purpose,
the coframe ${\vartheta}^{\alpha}$ is assumed to be of
Schwarzschild-deSitter (or Kottler) form,
\begin{equation}\label{coframe}
 {\vartheta}^{0}  =  e^{{\mu}(r)}dt\, , \quad
  {\vartheta}^{1}  =  e^{-{\mu}(r)}dr\, , \quad
  {\vartheta}^{2}  =  r\,d{\theta}   \, , \quad
  {\vartheta}^{3}  =  r\,{\sin} {\theta}\, d{\phi} \, ,
\end{equation}
with the function
\begin{equation}\label{Schwarz}
  e^{2{\mu}(r)} = 1 - 2\frac{m}{r} - \frac{{\lambda}_0}{3}r^2\,.
\end{equation}
We use Schwarzschild coordinates $x^i=(t,r,\theta,\phi)$.  Since the
coframe is assumed to be {\it orthonormal,} the metric reads
\begin{equation}\label{metric} g= -\vartheta ^{ {0}}\otimes
  \vartheta ^{ {0}}+ \vartheta ^{ {1}}\otimes \vartheta
  ^{ {1}}+ \vartheta ^{ {2}}\otimes \vartheta ^{ {2}}+
  \vartheta ^{ {3}}\otimes \vartheta ^{ {3}}\,.
\end{equation}

The nonmetricity $Q^{\alpha\beta}$ is given by
\begin{eqnarray}
Q^{\alpha\beta} & = &  \frac{{\ell}_{0}e^{-{\mu}(r)}}{2r^2}\pmatrix{
    {\vartheta}^{1} & 0 & 0 & 0 \cr
    0 & 0 & {\vartheta}^{2} & {\vartheta}^{3} \cr
    0 & {\vartheta}^{2} & 0 & 0 \cr
    0 & {\vartheta}^{3} & 0 & 0 }
 +
   \frac{{\ell}_{1}e^{-{\mu}(r)}}{2r^2}\pmatrix{
     3{\vartheta}^{0} & 0 & -{\vartheta}^{2} & -{\vartheta}^{3} \cr
     0 & -({\vartheta}^{0}-3{\vartheta}^{1} ) & 0 & 0 \cr
     -{\vartheta}^{2} & 0 & 0 & 0 \cr
     -{\vartheta}^{3} & 0 & 0 & 0 }\cr
& & \cr
& & \cr
&= &
   \frac{e^{-{\mu}(r)}}{2r^2}
   \pmatrix{
  3{\ell}_{1}{\vartheta}^{0}+{\ell}_{0}{\vartheta}^{1} & 0 &
      -{\ell}_{1}{\vartheta}^{2} &
      -{\ell}_{1}{\vartheta}^{3} \cr
  0 & -{\ell}_{1}({\vartheta}^{0}-3{\vartheta}^{1}) &
     {\ell}_{0}{\vartheta}^{2} &
                     {\ell}_{0}{\vartheta}^{3} \cr
  -{\ell}_{1}{\vartheta}^{2} & {\ell}_{0}{\vartheta}^{2} &
      0 & 0 \cr
  -{\ell}_{1}{\vartheta}^{3} & {\ell}_{0}{\vartheta}^{3} & 0 & 0 }\,.
\label{nonmetricity_O3nr2}
\end{eqnarray}
The integration constants ${\ell}_{0}$ and ${\ell}_{1}$ can be
interpreted as a measure for the violation of Lorentz invariance.
According to (\ref{nonmetricity_O3nr2}), we have $^{(2)}Q^{\a\b}=0$
or $P_\a=0$. All other irreducible pieces of $Q_{\a\b}$ are
nonvanishing: $^{(1)}Q^{\alpha\beta} \neq 0\, ,\,
^{(3)}Q^{\alpha\beta} \neq 0\, ,\,^{(4)}Q^{\alpha\beta} \neq 0\,$.
In particular, we find for the {\it shear\/} and the {\it Weyl\/}
1-forms
\begin{eqnarray}
\Lambda&=&  \frac{9 e^{-{\mu}(r)}}{8r^2}\,
                 ({\ell}_{0}+{\ell}_{1})\,{\vartheta}^{1}\,,\\
 Q & = &
 \frac{e^{-{\mu}(r)}}{8r^2}\,\left[-4\ell_1\,\vta^0
    +(3\ell_1-\ell_0)\,\vta^1\right]\,.
\end{eqnarray}
The torsion 2-form turns out to be
\begin{equation}
  T^{\alpha} = \ell_0 \frac{e^{-{\mu}(r)}}{4r^2}\pmatrix{
    {\vartheta}^{01} \cr 0 \cr -{\vartheta}^{12} \cr -{\vartheta}^{13}
    }\,-\ell_1 \frac{e^{-{\mu}(r)}}{4r^2}\pmatrix{ 0 \cr
    {\vartheta}^{01} \cr {\vartheta}^{02} \cr
    {\vartheta}^{03}}\,=\frac{e^{-{\mu}(r)}}{4r^2}\pmatrix{
    {\ell}_{0}{\vartheta}^{01} \cr -{\ell}_{1}{\vartheta}^{01} \cr
    -{\ell}_{1}{\vartheta}^{02}-{\ell}_{0}{\vartheta}^{12} \cr
    -{\ell}_{1}{\vartheta}^{03}-{\ell}_{0}{\vartheta}^{13} }\,.
\label{torsion_O3nr2}
\end{equation}
As a consequence, $^{(1)}T^{\alpha}=\,^{(3)}T^{\alpha}=0$ and only
$^{(2)}T^{\alpha}\neq 0$. By contraction of
(\ref{torsion_O3nr2}) with $e_\a\rfloor$ we find (recall
$T=e_\a\rfloor T^\a$)
\begin{equation}
  T = \frac{3e^{-{\mu}(r)}}{4r^2}\left( {\ell}_{1}{\vartheta}^{0}+
    {\ell}_{0}{\vartheta}^{1}\right)\,.
\end{equation}

The requirement that the torsion (\ref{torsion_O3nr2}) and the
nonmetricity (\ref{nonmetricity_O3nr2}) together with the
orthonormal coframe field (\ref{coframe}) be a solution of the field
equations of the Lagrangian (\ref{toy2}) implies some constraints on
the coupling constants. If we take care of these constraints in
(\ref{toy2}), we find the truncated Lagrangian
\begin{eqnarray}
V&=& \frac{1}{2\kappa}\,\left[  -R^{\alpha\beta}\wedge
      \eta_{\alpha\beta}-2\lambda_0\,\eta +
 Q_{\alpha\beta}\wedge{}^\star\Big(
        \frac{1}{4}\,^{(1)}Q^{\alpha\beta}
       -\frac{1}{2}\,^{(3)}Q^{\alpha\beta} \Big)\right]\nonumber\\
&&- \,\frac{z_{3}}{2\rho}\;^{(3)}Z^{\alpha\beta}
\wedge{}^\star\!\,^{(3)}Z_{\alpha\beta}\,. \label{nondeg1}
\end{eqnarray}
In other words, our exact solution
(\ref{coframe})-(\ref{nonmetricity_O3nr2}),(\ref{torsion_O3nr2})
solves the field equations of the Lagrangian (\ref{nondeg1}). Note
that $z_3$ is left arbitrary.

The explicit expressions for the strain and the rotational curvature
can be found in\footnote{According to the classification scheme of
  Baekler et al.\ \cite{MAGIII}, this solution is a special subcase of
  class Va.} Appendix \ref{cursol1}. The Weyl and the scalar parts of
the rotational curvature $W_{\a\b}$, see (\ref{F1}) and (\ref{F6}),
are composed of a Riemannian and a post-Riemannian piece. The other
parts of $W_{\a\b}$ as well as of $Z_{\a\b}$ are purely
post-Riemannian. We were surprised that $^{(3)}Z^{\alpha\beta}=0$;
particularly simple is {\tt dilcurv} $\sim \ell_1/(2r^3)$.

The set of the three one-forms $\{Q,T,{\Lambda}\}$ are related by
\begin{equation}
  3Q+2T-{\Lambda} = 0\, ,
\end{equation}
as can be checked easily. In this way, the torsion 1-form is closely
related to the shear and the Weyl 1-forms. This is a relation which
follows from Baekler's general prolongation ansatz
\cite{Baekler2003} to solve the field equations of MAG.

Our solution looks like a superposition of two elementary solutions.
For $\ell_1=0$ we find one elementary solution, a second one for
$\ell_0=0$, both for the same set of coupling constants. Also this fact
can be understood from the point of view of the prolongation
technique.

%******************************************************
\section{Killing vectors and quasilocal charges}\label{Killing}
%******************************************************

Let us determine the mass and the angular momentum of our exact
solution.  In a Riemannian space we call $\xi = \xi^\alpha \,
e_{\alpha}$ a Killing vector if the latter is the generator of a
symmetry transformation of the metric, i.e.,
\begin{equation}\label{killing1}
  {\pounds}_\xi \, g =0 \,.
\end{equation}
In metric-affine space, coframe and connection are
independent.  Hence, Eq.(\ref{killing1}) has to be supplemented by a
corresponding requirement for the connection \cite[p.83]{PRs},
\begin{equation}
  \pounds_\xi \, \Gamma_\alpha{}^\beta = 0 \,.
\end{equation}
These two relations can be recast into a more convenient form,
\begin{eqnarray}
  e_{(\alpha} \rfloor \widetilde D\xi_{\beta)} &=& 0 \,,\\
  D\left(e_\alpha \rfloor \stackrel{\frown}{D} \xi^\beta\right) +\xi
  \rfloor R_\alpha{}^\beta &=&0\,,
\end{eqnarray}
where $\widetilde D$ refers to the Riemannian part of the connection
(Levi-Civita connection) and $\stackrel{\frown}{D}$ to the transposed
connection: $\stackrel{\frown}{D}:=d +
\stackrel{\frown}{\Gamma}_\alpha{}^\beta :=d+ \Gamma_\alpha{}^\beta +
e_\alpha \rfloor T^\beta$.

For our solution the Killing vectors are the same as in case of the
Schwarzschild-de~Sitter metric in Riemannian spacetime,
\begin{eqnarray}
\stackrel{(0)}{\xi} &=&\hspace{10pt} \partial_t \,, \\
\stackrel{(1)}{\xi} &=&\hspace{10pt} \sin \phi \, \partial_\theta + \cot\theta \cos\phi
\, \partial_\phi \,,\\
\stackrel{(2)}{\xi} &=& -\cos \phi \, \partial_\theta + \cot\theta \sin\phi
\, \partial_\phi \,,\\
\stackrel{(3)}{\xi} &=& \hspace{10pt} \partial_\phi \,.
\end{eqnarray}
Subsequently we can compute the quasilocal charges by using formulas
of Nester, Chen, Tung, and Wu
\cite{Chen94,Ho97,Yu-Huei1,Yu-Huei2,Yu-Huei3}, for related work see
\cite{BaeklerShirafuji,Komar}, e.\,g.  The barred quantities refer to a
background solution, the symbol $\Delta$ denotes the difference
between a solution and the background, $\Delta\alpha = \alpha-\alpha$.
\begin{eqnarray}
{\mathfrak B}(N) &:=&
-
\left\{
\begin{array}{c}
\frac{1}{2} \, \Delta g_{\alpha\beta} \, \left(N \rfloor
\bar{M}^{\alpha\beta} \right) \\
\frac{1}{2} \, \Delta g_{\alpha\beta} \, \left(N \rfloor
{M}^{\alpha\beta} \right)
\end{array}
\right\}
-
\left\{
\begin{array}{rcl}
  (N\rfloor \vartheta^\alpha ) \, \Delta H_\alpha &+& \Delta
  \vartheta^\alpha \wedge \left(N \rfloor \overline{H}_\alpha\right)\\
  (N \rfloor \overline{\vartheta}^\alpha) \, \Delta H_\alpha &+&
  \Delta \vartheta^\alpha \wedge \left(N \rfloor H_\alpha \right)
\end{array}\right\} \nonumber \\
&& - \left\{\begin{array}{rcl} (\stackrel{\frown}{D}_\alpha N^\beta )
    \, \Delta H^{\alpha}{}_{\beta} &+& \Delta \Gamma_{\alpha}{}^{\beta}
    \wedge \left(N \rfloor \overline{H}^{\alpha}{}_{\beta} \right)
\vspace{5pt}\\
    \overline{(\stackrel{\frown}{D}_\alpha N^\beta)} \, \Delta
    H^{\alpha}{}_{\beta} &+& \Delta\Gamma_{\alpha}{}^{\beta} \wedge \left(N
      \rfloor H^{\alpha}{}_{\beta}\right)
\end{array}\right\}\,.
\end{eqnarray}
The upper (lower) line in the braces is chosen if the field strengths
(momenta) are prescribed on the boundary. By taking $N=\partial_t$ and
integrating ${\mathfrak B}$ over a 2-sphere and performing the limit
$r\to\infty$, we get the total energy. As background solution we
assume our solution with $m=0$ and $\ell_0=\ell_1=0$. Similarly, by
taking $N=\partial_\phi$ we obtain the total angular momentum,
\begin{eqnarray}
E_{\infty} &=& \lim_{r\to\infty}\int_{S^2} {\mathfrak B}(\partial_t) =
-\frac{8\pi \, m}{\kappa} \,,\\
L_{\infty}&=&\lim_{r\to\infty}\int_{S^2} {\mathfrak B}(\partial_\phi)=0 \,.
\end{eqnarray}

%%%%%%%%%%%%%%%%%%%%%%%%%%%%%%%%%%%%%%%%%%%%%%%%%%%%%%%%%%%%%%%%%%%%
\section{Remarks on the prolongation technique}\label{prolongation}
%%%%%%%%%%%%%%%%%%%%%%%%%%%%%%%%%%%%%%%%%%%%%%%%%%%%%%%%%%%%%%%%%%%%

Previously numerous exact solutions of MAG have already been found.
Let us mention, as examples, the papers \cite{Ho97,
  MinkevichVas03,YuriEffective, Romu1,Romu2,Romu3,
  PuetzfeldTresguerres, Puetzfeld2002,King,Dima1,Dima2,MAGII,
  Yu-Huei1,Yu-Huei2} and references given there. Even possible links
to observation were discussed in
\cite{Adak,PreussDr,Preuss1,Solanki,Puetzfeld2004a, Puetzfeld2004b}.
The solution in Sec.\ref{simplesolution} was found by using
prolongation methods. Such a method for MAG was proposed by Baekler et
al.\ \cite{Baekler2003,MAGIII}. In the sequel, we will explain how we
applied the prolongation method to our case in question.

For this purpose, we start, in the framework of the Poincar\'e gauge
theory (see \cite{Milutin,Erice95}), {}from a known exact solution
with Schwarzschild metric and $1/r^2$-torsion in a RC-spacetime, i.e.,
the nonmetricity vanishes.  Then, for {\it generating\/} nonmetricity,
we make the ansatz (Weyl 1-form $Q=Q_\alpha{}^\a/4$)
\begin{eqnarray}\label{ansatz}
  T^\alpha & =& \xi_0\,Q^\alpha{}_\beta\wedge \vartheta^\beta +
  \xi_1\,Q\wedge \vartheta^\alpha +\, ^{(3)}T^{\alpha}\nonumber \\ &
  =& {\xi}_{0}\,\qslash ^{\alpha}{}_{\beta}\wedge {\vartheta}^{\beta}
  +( {\xi}_{0}+ {\xi}_{1})\,Q\wedge {\vartheta}^{\alpha} +
  {^{(3)}T^{\alpha}}\,,\\ 24 & = &\hspace{30pt} 16\hspace{30pt}
  {+}\hspace{30pt} 4\hspace{30pt} {+}\hspace{30pt} 4
  \nonumber\,,
\end{eqnarray}
with arbitrary constants ${\xi}_{0}$ and ${\xi}_{1}$ to be determined
by the field equations. The 24 components of $T^\a$ are related to the
40 components of $Q_{\a\b}$.  Because of the property
$^{(1)}Q^{\alpha}{}_{\beta}\wedge {\vartheta}^{\beta}=0$, the
irreducible part $^{(1)}Q^{\alpha}{}_{\beta}$ (16 independent
components) does not contribute to the torsion. Furthermore
$(Q^\a{}_\beta\wedge {\vartheta}^{\beta})\wedge \vta_\a=0$ (4
independent components). Hence, our ansatz relates the 24 components
of the torsion to $16+4+4 = 40-16$ independent components of
$Q^{\alpha}{}_{\beta}\wedge {\vartheta}^{\beta}$, $Q\wedge
{\vartheta}^{\alpha}$, and $^{(3)}T^{\alpha}$.  In the end, the second
order partial differential equations (\ref{zeroth'}),(\ref{second'})
become nonlinear algebraic equations.  We solve them by requiring
certain {\it constraints\/} on the coupling constants. In this way we
find exact solutions with $1/r^\nu$-behavior of $Q_{\a\b}$, here
$\nu=1,2,3$.

If we substitute the decomposition formula (\ref{qwetchtheta}) into
(\ref{ansatz}), we find
\begin{equation}\label{ansatz'}
  T^{\alpha}=\xi_0\,P^\a- \frac{1}{3}\,\left[\xi_0\,\Lambda-3
    (\xi_0+{\xi}_{1})\,Q\right]\wedge {\vartheta}^{\alpha}+
  {^{(3)}T^{\alpha}}\,.
\end{equation}
Contraction yields
\begin{equation}\label{ansatz''}
  \xi_0\,\Lambda-3(\xi_0+\xi_1)\,Q-T=0\,.
\end{equation}
Here it is useful to take recourse to the first Bianchi identity.
Provided that the conditions
\begin{equation}\label{tracepiece'}
T^\alpha = {}^{(2)}T=\frac{1}{3} \, \vta^\alpha \wedge T \,,
\quad {}^{(2)}Q_{\alpha\beta}=0\,,
\end{equation}
are fulfilled, we derive in Appendix \ref{firstBia} in
(\ref{integrab}) that [$\ra:=\vta^\alpha \wedge (e_\beta\rfloor
\,^{(5)}W_\alpha{}^\beta)$]
\begin{equation}\label{integrab'}
  \ra - \frac{1}{3} \, d \left(2T + 3 Q -\Lambda\right) =0 \,.
\end{equation}
This equation can be understood as an integrability condition for
(\ref{ansatz''}). Substitution of (\ref{ansatz''}) into
(\ref{integrab'}) yields
\begin{equation}\label{combined}
  0 = \ra + \frac{1}{3}d\left\{ (2{\xi}_{0}-1){\Lambda} +3
    \left[1 -2({\xi}_{0}+{\xi}_{1})\right] Q \right\}\,.
\end{equation}
We have two free parameters. Thus, on this level, we can always find
solutions with $\ra=0$. However, Eq.(\ref{combined}) also
implies the non-trivial result
\begin{equation}
  d\,{\ra}=0\,.
\end{equation}
The second Bianchi identity may yield more conditions.

We now turn to the spherically symmetric solution of
Sec.\ref{simplesolution}. In this case, we have the prolongation
ansatz (\ref{ansatz}) with
\begin{equation}\label{xx}
  \xi_0=\frac 12\,,\quad \xi_1=0\,,\quad \,^{(3)}T^\a=0\,.
\end{equation}
Then (\ref{ansatz''}) becomes
\begin{equation}
  {\Lambda} -3Q - 2T = 0 \,.
\end{equation}
This is consistent with the first Bianchi identity, see
(\ref{letzte}). Since only $^{(2)}T^\a\ne 0$, we have
\begin{equation}
  {\Lambda} -3Q - 2T = {\rm exact\; form}\,.
\end{equation}
The distortion 1-form (\ref{N}) can be taken {}from
(\ref{nonmetricity_O3nr2}) and (\ref{torsion_O3nr2}) or {}from
\cite{MAGIII}, Eq.(35):
\begin{equation}
  N_{\alpha\b}=\frac 12\,Q_{\a\b}\,.
\end{equation}
Note that $N_{[\alpha\b]}=0$. In this special case, the curvature can
be easily decomposed in Riemannian and post-Riemannian pieces,
\begin{equation}\label{first_bianchi}
  {R_{\alpha}}^{\beta}= \widetilde{R}_{\alpha}{}^{\beta} + {1\over 2}
  \widetilde{D}{Q_{\alpha}}^{\beta}-{1\over 4}
  {\qslash_{\alpha}}^{\gamma}\wedge {\qslash_{\gamma}}^{\beta} \,,
\end{equation}
where $\widetilde{R}_{\alpha}{}^{\beta}$ denotes the purely Riemannian
part of the curvature and $\widetilde{D}$ the exterior covariant
derivative with respect to the Riemannian connection.

Possibly, for a real ``liberated'' aether dynamics, one is forced to
allow for $\ra\ne 0$. Apparently $\ra$ is the non-exact piece of
$\Lambda,Q,T$ and as such contributes generically to $^{(3)}Z_{\a\b}$
and $^{(4)}Z_{\a\b}$.

There is a further property of our specific exact solution which is of
interest. The ansatz (\ref{ansatz}), together with the first Bianchi
identity, yields
\begin{eqnarray}\label{G2}
 \left(Q_{\alpha\mu}\wedge Q_{\beta}\,^{\mu}+ 4R_{[\alpha\beta ]} \right)
\wedge {\vartheta}^{\beta}& =&\cr \left(\qslash_{\alpha\mu}\wedge
\qslash_{\beta}\,^{\mu}+ 4R_{[\alpha\beta ]} \right) \wedge
{\vartheta}^{\beta}& =& 0\,.
\end{eqnarray}

Compare now (\ref{first_bianchi}) with (\ref{G2}) and find
\begin{equation}
  R_{\alpha\beta}\wedge{\vartheta}^{\beta} = {\widetilde
    R}_{\alpha\beta}\wedge {\vartheta}^{\beta}+W_{\alpha\beta}\wedge
  {\vartheta}^{\beta}+ \frac{1}{2}( {\widetilde
      D}Q_{\alpha\beta})\wedge {\vartheta}^{\beta}\,,
\end{equation}
i.e., we have ${\xi}_{0}=1/2$ and ${\xi}_{1}=0$. This
implies for the symmetric (strain) curvature
\begin{equation}
  Z_{\alpha\beta}\wedge {\vartheta}^{\beta} = {\widetilde
    R}_{\alpha\beta}\wedge {\vartheta}^{\beta} + \frac{1}{2}(
    {\widetilde D}Q_{\alpha\beta})\wedge {\vartheta}^{\beta}\,.
\end{equation}
We decompose the curvature into symmetric and antisymmetric pieces.
This yields%, together with the properties (\ref{WWW}),
\begin{eqnarray}
  Z_{\alpha\beta} & = & \frac{1}{2}( {\widetilde
      D}Q_{\alpha\beta})\,, \\ \nonumber & & \\ W_{\alpha\beta}
  & = & {\widetilde R}_{\alpha\beta} - \frac{1}{4}{\qslash}_{[\alpha
    }{}^{\gamma}\wedge {\qslash}_{\beta ]\gamma}\,.
\end{eqnarray}
Thus an extra field equation for ${\qslash}_{\alpha\beta}$ is implied,
\begin{equation}
  ( {\widetilde D}{\qslash}_{\alpha\beta})\wedge
  {\vartheta}^{\beta} = 0\,.
\end{equation}
Generally, this equation implies further integrability conditions.
With
\begin{equation}
  DD{\qslash}_{\alpha\beta} = -2R_{( \alpha }{}^{\g}\wedge
    {\qslash}_{\beta )\g}
\end{equation}
and the ansatz (\ref{ansatz}), we find the algebraic constraint
\begin{equation}
  \frac{1}{{\xi}_{0}}{\widetilde R}_{\alpha\b}\wedge
  T^{\b}=0\,,
\end{equation}
which has to be fulfilled by this solution.

%%%%%%%%%%%%%%%%%%%%%%%%%%%%%%%%%%%%%%%%%%%%%%%%%%%%%%%%%%%%%%%%%%%%
\section{Discussion}\label{discussion}
%%%%%%%%%%%%%%%%%%%%%%%%%%%%%%%%%%%%%%%%%%%%%%%%%%%%%%%%%%%%%%%%%%%%

We constructed a model within MAG which exhibits vector-like Lorentz
violating fields. It may be seen as analogue to the Einstein-aether
theory of Jacobson et al. We were able to find a simple exact
spherically symmetric solution for the field equations of a truncated
Lagrangian. Our solution distorts the conventional
spherically symmetric Schwarzschild-de Sitter spacetime by
nonmetricity and torsion. The presence of nonmetricity will obstruct
the local Lorentz invariance of the Riemannian spacetime.  Further
investigations should include the search for wave-like aether solution
which, most likely, will require a more complicated ``background''
than simple Schwarzschild spacetime, cf.\ plane-wave solutions in MAG
\cite{MAGplane,Macias00,Yurippwaves}.

It should be understood that MAG is a comprehensive framework for
classical gravitational field theories.  Different authors started
with MAG and, by using the nonlinear realization technique, tried to
``freeze out'' certain degrees of freedom like, e.g., the
nonmetricity. Percacci \cite{Percacci}, Tresguerres and Mielke
\cite{Romu3} (see also \cite{Romu4,Romu5}), and, most recently, Kirsch
\cite{ingo} developed models of such a kind, for earlier work one
should compare Lord and Goswami \cite{Lord1987,Lord+Goswami}, see also
Tresguerres \cite{Romu6}. There is not much doubt that gravity has a
metric-affine structure. Therefore, MAG seems an appropriate framework
for classical gravity. But there are different ways to realize this
structure. Still, if Lorentz invariance turns out to be violated, then
the nonmetricity of spacetime should play a leading role.

\begin{acknowledgments}
  We thank Robert Bluhm, Ted Jacobson, Dirk Puetzfeld, Yakov Itin,
  Ingo Kirsch, and Romualdo Tresguerres for valuable comments on the
  manuscript.  This project has been supported by the grant HE
  528/20-1 of the DFG, Bonn.
\end{acknowledgments}

\appendix

%%%%%%%%%%%%%%%%%%%%%%%%%%%%%%%%%%%%%%%%%%%%%%%%%%%%%%%%%%%%%%%%%%%%
\section{Irreducible decomposition of the strain
  curvature $Z_{\a\b}$ in $n$ dimensions}\label{straincurv}
%%%%%%%%%%%%%%%%%%%%%%%%%%%%%%%%%%%%%%%%%%%%%%%%%%%%%%%%%%%%%%%%%%%%

The 2-form $\zslash_{\a\b}$ in (\ref{Zslash}) --- for $n$ dimensions
we have $1/n$ instead of $1/4$ --- can be cut into different pieces by
contraction with $e_\a$, transvecting with $\vta^\a$, and by
``hodge''-ing the corresponding expressions:
\begin{equation}\label{ZDY}\zslash_\a:=e^\b\rfloor \zslash_{\a\b}, \qquad
  \hat\Delta:={1\over n-2}\,\vta^\a\wedge\zslash_\a,\qquad
  Y_\a:=\,^*(\zslash_{\a\b}\wedge\vta^\b)\,.\end{equation}
Subsequently we can subtract out traces:
\begin{equation}\Xi_\a:=
  \zslash_\a-{1\over2}e_\a\rfloor(\vta^\g\wedge\zslash_\g),
  \qquad\qquad \Upsilon_\a:= Y_\a- {1\over
    n-2}\,e_\a\rfloor(\vta^\g\wedge Y_\g)\,.\end{equation}
The irreducible pieces may then be written as (see \cite{PRs})
\begin{eqnarray}
^{(2)}\!Z_{\a\b}&:=& - {1\over 2}\,^*(\vta_{(\a}
\wedge\Upsilon_{\b)})\,,\\
\label{Z3de}^{(3)}\!Z_{\a\b}&:=&\hspace{8pt}{1\over n+2}\;\left(
n\,\vta_{(\a}\wedge e_{\b)}\rfloor-2\,g_{\a\b}\right)\hat\Delta\,,\\
^{(4)}\! Z_{\a\b}&:=&\hspace{8pt} {1\over n}\;g_{\a\b}\,Z\,,\\
^{(5)}\! Z_{\a\b}&:=&\hspace{8pt}{2\over n}\; \vta_{(\a}\wedge
\Xi_{\b)}\,,\\
^{(1)}\!Z_{\a\b}&:=&\hspace{8pt} Z_{\a\b}-
\,^{(2)}\!Z_{\a\b}-\,^{(3)}\!Z_{\a\b}-\,^{(4)}\!Z_{\a\b}-
\,^{(5)}\!Z_{\a\b}\,.\end{eqnarray}

%%%%%%%%%%%%%%%%%%%%%%%%%%%%%%%%%%%%%%%%%%%%%%%%%%%%%%%%%%%%%%%%%%%%
\section{Expressing the curvature $^{(3)}Z_{\a\b}$ in terms of
  nonmetricity and torsion}\label{hatDelta}
%%%%%%%%%%%%%%%%%%%%%%%%%%%%%%%%%%%%%%%%%%%%%%%%%%%%%%%%%%%%%%%%%%%%

According to (\ref{Z3de}), the curvature $^{(3)}Z_{\a\b}$ can be
expressed in terms of $\hat\Delta$. Thus, we start {}from the
definition of $\hat\Delta$ in (\ref{ZDY})$_2$ and move the interior
product to the left:
\begin{equation}
\hat\Delta = \frac{1}{n-2} \, \vta^\alpha \wedge
\left(e^\beta \rfloor \zslash_{\alpha\beta}\right)
= \frac{1}{n-2} \, \left[ -e^\beta \rfloor (\vta^\alpha \wedge
\zslash_{\alpha\beta} ) + \zslash_\alpha{}^\alpha \right]
= -\frac{1}{n-2} e^\alpha \rfloor (\zslash_{\alpha\beta}\wedge \vta^\beta)
\,.
\end{equation}
Obviously, we have to express $\zslash_{\alpha\beta}\wedge \vta^\beta$
in terms of nonmetricity and torsion. This should be possible by means
of the zeroth Bianchi identity
\begin{equation}\label{0bia1}
  DQ_{\alpha\beta}= -DD g_{\alpha\beta} = R_\alpha{}^\gamma \,
g_{\gamma\beta} + R_\beta{}^\gamma \, g_{\alpha\gamma}
=
2 R_{(\alpha\beta)} = 2 Z_{\alpha\beta}\,.
\end{equation}
We wedge with $\vta^\b$ {}from the right and obtain
\begin{equation}\label{0bia2}
  D\left(Q_{\alpha\beta}\wedge\vartheta^\beta\right) =
  2Z_{\alpha\beta} \wedge\vartheta^\beta - Q_{\alpha\beta} \wedge
  T^\beta \,.
\end{equation}
On the other hand, by making use of (\ref{qwetchtheta}) (suitably
generalized for $n$ dimensions) and of
\begin{equation}\label{Do_a}
D\vartheta_\alpha = D(g_{\alpha\beta} \, \vartheta^\beta )
= (Dg_{\alpha\beta}) \wedge \vartheta^\beta + g_{\alpha\beta} \,
D\vartheta^\beta = -Q_{\alpha\beta} \wedge \vartheta^\beta+ T_{\alpha} \,,
\end{equation}
we can calculate
\begin{eqnarray}
D\left(Q_{\alpha\beta}\wedge\vartheta^\beta\right)
&\stackrel{(\ref{qwetchtheta})}{=}&
D\left(Q\wedge\vartheta_\alpha +\frac{1}{n-1} \, \vartheta_\alpha \wedge
\Lambda+ P _\alpha\right) \nonumber \\
&=&
dQ \wedge \vartheta_\alpha - Q\wedge D\vartheta_\alpha
+\frac{1}{n-1} \, D\vartheta_\alpha \wedge \Lambda
- \frac{1}{n-1}\vartheta_\alpha \wedge d \Lambda \nonumber
+ D P _\alpha\\
&\stackrel{(\ref{Do_a})}{=}&
dQ \wedge \vartheta_\alpha
- Q\wedge \left(-Q_{\alpha\beta}\wedge \vartheta^\beta + T_\alpha \right)
\nonumber \\&&
+ \frac{1}{n-1} \, \left( -Q_{\alpha\beta}\wedge \vartheta^\beta + T_\alpha
   \right) \wedge \Lambda
- \frac{1}{n-1} \, \vartheta_\alpha \wedge d \Lambda
+ D  P _\alpha \nonumber \\
&\stackrel{(\ref{qwetchtheta})}{=}&
dQ \wedge \vartheta_\alpha
+ Q\wedge \left( Q\wedge \vartheta_\alpha + \frac{1}{n-1} \,
\vartheta_\alpha \wedge \Lambda +  P _\beta\right)
- Q\wedge T_\alpha
\nonumber \\&&
 -\frac{1}{n-1} \, \left( Q\wedge
\vartheta_\alpha + \frac{1}{n-1} \, \vartheta_\alpha \wedge \Lambda +
 P _\alpha \right) \wedge \Lambda
+\frac{1}{n-1} \, T_\alpha \wedge \Lambda
\nonumber \\ &&
- \frac{1}{n-1} \, \vartheta_\alpha \wedge d \Lambda
+ D P _{\alpha} \nonumber \\
&=&
  dQ \wedge \vartheta_\alpha
+ Q\wedge Q \wedge \vta_\alpha
+ \frac{1}{n-1} \, Q \wedge \vta_\alpha \wedge \Lambda
+ Q \wedge  P _\alpha
- Q \wedge T_\alpha
\nonumber \\ &&
- \frac{1}{n-1} Q \wedge \vta_\alpha \wedge \Lambda
- \frac{1}{(n-1)^2} \, \vta_\alpha \wedge \Lambda \wedge \Lambda
- \frac{1}{n-1} \,  P _\alpha \wedge \Lambda
+ \frac{1}{n-1} \, T_\alpha \wedge \Lambda
\nonumber \\ &&
- \frac{1}{n-1} \, \vta_\alpha \wedge d\Lambda
+ D P _\alpha
\nonumber \\ &=&
dQ \wedge \vta_\alpha
+ Q \wedge  P _\alpha
- Q \wedge T_\alpha
- \frac{1}{n-1} \,  P _\alpha \wedge \Lambda
+ \frac{1}{n-1} \, T_\alpha \wedge \Lambda
\nonumber \\ &&
- \frac{1}{n-1} \, \vta_\alpha \wedge d\Lambda
+ D  P _\alpha \,.\label{110}
\end{eqnarray}

Now we can compare (\ref{0bia2}) and (\ref{110}). We find
\begin{eqnarray}
2 \zslash_{\alpha\beta}\wedge \vta^\beta
&=&
2Z_{\alpha\beta} \wedge \vta^\beta -
2{}^{(4)}Z_{\alpha\beta}\wedge\vta^\beta \nonumber \\
&=&
  Q\wedge P_\alpha
- Q\wedge T_\alpha
- \frac{1}{n-1} \, P_\alpha \wedge \Lambda
\nonumber \\ && \label{zwedgevta}
+ \frac{1}{n-1} \, T_\alpha \wedge \Lambda
- \frac{1}{n-1} \, \vta_\alpha \wedge d\Lambda
+ DP_\alpha
+ Q_{\alpha\beta} \wedge T^\beta \,.
\end{eqnarray}
We expand the last term by means of the irreducible decomposition of
torsion and nonmetricity:
\begin{eqnarray}\hspace{-0.5cm}
Q_{\alpha\beta} \wedge T^\beta
&=&
\qslash_{\alpha\beta} \wedge T^\beta + Q\wedge T_{\alpha} \nonumber \\
&=&
\qslash_{\alpha\beta} \wedge \left( {}^{(1)}T^\beta + {}^{(3)}T^\beta \right)
+ \qslash_{\alpha\beta}\wedge \left(\frac{1}{n-1} \, \vta^\beta \wedge T
\right)
+ Q\wedge T_{\alpha} \nonumber \\ \label{qwedget}
&=&
\qslash_{\alpha\beta} \wedge \left( {}^{(1)}T^\beta + {}^{(3)}T^\beta \right)
+\frac{1}{(n-1)^2} \, \vta_\alpha \wedge \Lambda \wedge T + \frac{1}{n-1} \,
P_\alpha \wedge T + Q\wedge T_\alpha\,.
\end{eqnarray}
By the same token,
\begin{equation}\label{twedgelam}
  \frac{1}{n-1} \, T_\alpha \wedge \Lambda = \frac{1}{n-1} \,
  \left({}^{(1)}T_\alpha + {}^{(3)}T_\alpha\right)\wedge\Lambda +
  \frac{1}{(n-1)^2} \, \vta_\alpha \wedge T \wedge \Lambda \,.
\end{equation}
Substituting (\ref{qwedget},\ref{twedgelam}) into (\ref{zwedgevta}) yields
\begin{eqnarray}
2 \zslash_{\alpha\beta}\wedge \vta^\beta
&=&
DP_\alpha - \frac{1}{n-1} \, \vta_\alpha \wedge d\Lambda +
\qslash_{\alpha\beta} \wedge \left({}^{(1)}T^\beta + {}^{(3)}T^{\beta}\right)
\nonumber \\ &&
+ \frac{1}{n-1} \, \left({}^{(1)}T_\alpha + {}^{(3)}T_\alpha\right) \wedge
\Lambda + P_\alpha \wedge \left[Q-\frac{1}{n-1} \, (\Lambda - T)\right]
\,.
\end{eqnarray}
We use the following properties of the irreducible pieces:
\begin{equation}
e^\alpha \rfloor P_\alpha = e^\alpha \rfloor {}^{(1)}T_\alpha
= e^\alpha \rfloor {}^{(3)}T_\alpha = 0\,,\qquad
e^\alpha \rfloor \qslash_{\alpha\beta} = \Lambda_\beta \,.
\end{equation}
Then we find
\begin{eqnarray}
2e^\alpha \rfloor (\zslash_{\alpha\beta}\wedge \vta^\beta)
&=&
P^\alpha \, e_\alpha \rfloor \left[ Q-\frac{1}{n-1} \, (\Lambda-T)\right]
+ \frac{n}{n-1} \, \left({}^{(1)}T_\alpha + {}^{(3)}T_\alpha \right) \,
\, \Lambda^\alpha \nonumber \\ &&
- \frac{n-2}{n-1} \, d\Lambda + e^\alpha \rfloor DP_\alpha
- \qslash_{\alpha\beta} \wedge e^\alpha\rfloor \left({}^{(1)}T^\beta +
{}^{(3)}T^\beta \right) \,.
\end{eqnarray}
We can further simplify the last term. First we note that
\begin{equation}
e^\alpha \rfloor {}^{(3)}T^\beta
=
(-1)^s \, e^\alpha \rfloor
\frac{1}{3} \, {}^\star\left[ \vta^\beta \wedge {}^\star (T^\gamma
\wedge \vta_\gamma)  \right]
=
(-1)^s \, \frac{1}{3} \, {}^\star\left[ \vta^\beta \wedge {}^\star (T^\gamma
\wedge \vta_\gamma)  \wedge \vta^\alpha \right]
=
- e^\beta \rfloor {}^{(3)}T^\alpha \,.
\end{equation}
Hence,
\begin{eqnarray}
\qslash_{\alpha\beta} \wedge e^\alpha\rfloor \left({}^{(1)}T^\beta +
{}^{(3)}T^\beta \right) &=& \qslash_{\alpha\beta} \wedge e^\alpha\rfloor
{}^{(1)}T^\beta
=
  {}^{(1)}Q_{\alpha\beta} \wedge e^\alpha \rfloor {}^{(1)}T^\beta
\nonumber \\ &&
+ {}^{(2)}Q_{\alpha\beta} \wedge e^\alpha \rfloor {}^{(1)}T^\beta
+ {}^{(3)}Q_{\alpha\beta} \wedge e^\alpha \rfloor {}^{(1)}T^\beta\,.
\end{eqnarray}
The last term can be further rewritten as
\begin{eqnarray}
{}^{(3)}Q_{\alpha\beta} \wedge e^\alpha \rfloor {}^{(1)}T^\beta
&=&
\frac{2n}{(n-1)(n+2)} \, \left(\vta_{(\alpha} \, \Lambda_{\beta)}
-\frac{1}{n} \, g_{\alpha\beta} \, \Lambda \right)
\wedge e^\alpha \rfloor {}^{(1)}T^\beta \nonumber \\
&=&
\frac{n}{(n-1)(n+2)} \, \left(
\vta_\beta \, \Lambda_\alpha \wedge e^\alpha\rfloor {}^{(1)}T^\beta
+ \Lambda_\beta \vta_\alpha  e^\alpha \rfloor {}^{(1)}T^\beta \right)
\nonumber \\
&=&
\frac{n}{(n-1)(n+2)} \, \left[\Lambda_\alpha \, \left(-e^\alpha
\rfloor(\vta_\beta \wedge {}^{(1)}T^\beta ) + {}^{(1)}T^\alpha \right)
+ 2 \Lambda_\beta \, {}^{(1)}T^\beta \right]
\nonumber \\
&=&
\frac{3n}{(n-1)(n+2)} \, \Lambda_\alpha \, {}^{(1)}T^\alpha \,,
\end{eqnarray}
where we used $\vta_\beta \wedge {}^{(1)}T^\beta=0$.
Finally we arrive at
\begin{eqnarray}\label{B15}
\hat\Delta &=&
\frac{1}{2(n-1)} \, d \Lambda - \frac{1}{2(n-2)} \, e^\alpha \rfloor
DP_\alpha \nonumber \\
&&
-\frac{1}{2(n-2)} \, \Bigg\{ \frac{1}{n-1} \, P_\alpha \; e^\alpha \rfloor
\left[(n-1) \, Q+ \Lambda - T \right]
+ \left(\frac{n+1}{n+2} \, {}^{(1)} T_\alpha +\frac{n}{n-1} \, {}^{(3)}
T_\alpha \right) \, \Lambda^\alpha
\nonumber \\ && \null \hspace{4cm}
-\left({}^{(1)}Q_{\alpha\beta}+{}^{(2)}Q_{\alpha\beta} \right) \wedge
e^\alpha \rfloor {}^{(1)}T^\beta\Bigg\} \,.
\end{eqnarray}
Note that in the last line we could substitute $^{(2)}Q_{\alpha\beta}
= -{2}\, e_{(\alpha} \rfloor P _{\beta)}/3$.

%%%%%%%%%%%%%%%%%%%%%%%%%%%%%%%%%%%%%%%%%%%%%%%%%%%%%%%%%%%%%%%%
\section{The $^{(3)}Z^{\alpha\beta} \wedge {}^{\star \,
    (3)}Z_{\alpha\beta}$ piece of the Lagrangian}\label{3Z2}
%%%%%%%%%%%%%%%%%%%%%%%%%%%%%%%%%%%%%%%%%%%%%%%%%%%%%%%%%%%%%%%%

We proof that the following relation holds for {\em arbitrary}
spacetimes:
\begin{equation}\label{C1}
  {}^{(3)}Z^{\alpha\beta} \wedge {}^{\star \, (3)}Z_{\alpha\beta} =
  \frac{n(n-2)}{n+2} \, \hat{\Delta} \wedge {}^\star \hat{\Delta}\,.
\end{equation}
This comes about since ${}^{(3)}Z^{\alpha\beta}$ corresponds to a {\em
  scalar}-valued degree of freedom, namely to the two-form
$\hat{\Delta}$, see (\ref{Z3de}). For a p-form $\phi$, we have the
rules for the Hodge dual $^{\star\star}\phi=(-1)^{p(n-p)-1}\phi\;$ in
the case of Lorentz signature, furthermore, $\vta^\alpha \wedge
(e_\alpha \rfloor \phi) = p \, \phi\;$ and
$^\star(\phi\wedge\vta_\a)=e_\a\rfloor\,^\star\phi\,$.  Thus, the
terms quadratic in contractions of $\hat{\Delta}$ in the end evaluate
to $\hat{\Delta} \wedge {}^\star \hat{\Delta}\,$,
\begin{equation}
  (e^\alpha \rfloor \hat{\Delta}) \wedge {}^\star (e_\alpha \rfloor
  \hat{\Delta}) =- (e^\alpha \rfloor \hat{\Delta}) \wedge {}^\star
  (e_\alpha \rfloor\, ^{\star\star}\hat{\Delta})=- (e^\alpha \rfloor
  \hat{\Delta}) \wedge \vta_\alpha \wedge {}^\star \hat{\Delta} = 2 \,
  \hat{\Delta} \wedge {}^\star \hat{\Delta}\,.
\end{equation}
We recall the definition (\ref{Z3de})
\begin{equation}\label{C3}
  {}^{(3)}Z_{\alpha\beta} = \frac{1}{n+2} \, \left[ n \,
    \vta_{(\alpha} \wedge e_{\beta)} \rfloor \hat{\Delta} - 2 \,
    g_{\alpha\beta} \, \hat{\Delta} \right]\,.
\end{equation}
It is symmetric $^{(3)}Z^{[\alpha\beta]}=0$ and tracefree $^{(3)}
Z^\g{}_\g=0$.
Thus,
\begin{eqnarray}
  {}^{(3)}Z_{\alpha\beta} \wedge {}^{\star} \, {}^{(3)}Z^{\alpha\beta}
  &=& \frac{1}{n+2} \, \left[ n \, \vta_{\alpha} \wedge (e_{\beta}
    \rfloor \hat{\Delta})\wedge {}^{\star} {}^{(3)}Z^{\alpha
      \beta}\right] \nonumber \\ &=& \frac{1}{(n+2)^2} \, \left[
    n^2\,\vta_{\alpha} \wedge (e_{\beta} \rfloor \hat{\Delta}) \wedge
    {}^{\star}\left( \vta^{( \alpha} \wedge (e^{\beta) } \rfloor
      \hat{\Delta})\right) - 2n \, \vta_{\alpha} \wedge (e_{\beta}
    \rfloor \hat{\Delta}) \wedge g^{\alpha\beta} \,
    {}^{\star}\hat{\Delta} \right]\,. \nonumber \\ \label{hallo}
\end{eqnarray}
In order to calculate the first term, we apply the rules for commuting the
Hodge star with the exterior/interior product.
\begin{eqnarray}
  {}^\star \left(\vta^{(\alpha} \wedge (e^{\beta)} \rfloor
    \hat{\Delta})\right) &=& -\,^\star\!\left[(e^{(\beta} \rfloor
    \hat{\Delta}) \wedge \vta^{\alpha)} \right] = - \, e^{(\alpha}
  \rfloor \,^{\star}\!\left( e^{\beta)} \rfloor \hat{\Delta} \right)
  \nonumber \\ &=& -(-1)^{n-1} \, e^{(\alpha} \rfloor \left( {}^\star
    \hat{\Delta} \wedge \vta^{\beta)} \right) = e^{(\alpha} \rfloor
  \left(\vta^{\beta)} \wedge {}^\star \hat{\Delta} \right) \nonumber
  \\ &=& g^{\alpha\beta} \, {}^\star \hat{\Delta} - \vta^{(\alpha}
  \wedge( e^{\beta)} \rfloor \,^\star \hat{\Delta}) \,.\label{hallo1}
\end{eqnarray}
The second term in the brackets of (\ref{hallo}) simply evaluates to
\begin{equation}
 - 2n \, \vta_{\alpha} \wedge (e_{\beta} \rfloor \hat{\Delta}) \wedge
  g^{\alpha\beta} \, {}^{\star}\hat{\Delta} = -2n \, \vta^\alpha
  \wedge (e_\alpha \rfloor \hat{\Delta}) \wedge {}^\star \hat{\Delta} =
  -4n \, \hat{\Delta} \wedge {}^\star \hat{\Delta}\,.\label{hallo2}
\end{equation}
Substituting (\ref{hallo1}) and (\ref{hallo2}) into (\ref{hallo}), we
find
\begin{equation}
  {}^{(3)}Z^{\alpha\beta} \wedge {}^{\star} \, {}^{(3)}Z^{\alpha\beta}
  = \frac{1}{(n+2)^2} \,\left[2n^2\hat{\Delta}\wedge\,
    ^\star\hat{\Delta}- n^2\,\vta_{\alpha} \wedge (e_{\beta} \rfloor
    \hat{\Delta}) \wedge \vta^{(\alpha} \wedge( e^{\beta)} \rfloor
    \,^\star \hat{\Delta}) -4n \, \hat{\Delta} \wedge {}^\star
    \hat{\Delta} \right]\,.\label{hallo4}
\end{equation}
The term in the middle yields ($\vta^\a\wedge\vta_\a=0$)
\begin{eqnarray}
  (\vta_{\alpha} \wedge e_{\beta} \rfloor \hat{\Delta} ) \wedge
  (\vta^{(\alpha} \wedge e^{\beta)} \rfloor {}^\star \hat{\Delta} )
  &=& \frac{1}{2} \, \left[ \vta_{\alpha} \wedge (e_{\beta} \rfloor
    \hat{\Delta} ) \wedge \vta^{\alpha} \wedge (e^{\beta} \rfloor
    {}^\star \hat{\Delta} ) + \vta_{\a} \wedge (e_{\b} \rfloor
    \hat{\Delta} ) \wedge \vta^{\b} \wedge (e^{\a} \rfloor {}^\star
    \hat{\Delta} )\right] \nonumber \\ &=&\frac{1}{2}\left[-\vta^{\b}
    \wedge (e_{\b} \rfloor \hat{\Delta} ) \wedge \vta_{\a} \wedge
    (e^{\a} \rfloor {}^\star \hat{\Delta} ) \right]= -(n-2) \,
  \hat{\Delta} \wedge {}^\star \hat{\Delta} \,.
\end{eqnarray}
Eventually,
\begin{equation}
  {}^{(3)}Z^{\alpha\beta} \wedge {}^{\star} \, {}^{(3)}Z^{\alpha\beta}
  = \frac{1}{(n+2)^2} \, \left[2n^2 + n^2(n-2)-4n\right] \,
  \hat{\Delta} \wedge {}^\star \hat{\Delta} = \frac{n(n-2)}{n+2} \,
  \hat{\Delta} \wedge {}^\star \hat{\Delta} \,.
\end{equation}

%%%%%%%%%%%%%%%%%%%%%%%%%%%%%%%%%%%%%%%%%%%%%%%%%%%%%%%%%%%%%%%%%%%%%%
\section{The MAG Lagrangian quadratic in curvature,
  torsion, and nonmetricity}\label{MAGLagrangian}
%%%%%%%%%%%%%%%%%%%%%%%%%%%%%%%%%%%%%%%%%%%%%%%%%%%%%%%%%%%%%%%%%%%%%%

The most general parity conserving quadratic Lagrangian, which is
expressed in terms of the $4+3+6+5$ irreducible pieces of
$Q_{\alpha\beta}$, $T^\alpha$, $W_\alpha{}^\beta$, and
$Z_\alpha{}^\beta$, respectively, reads (see \cite{Esser},
\cite{YuriEffective}, \cite{MAGII}, and references given):
\begin{eqnarray}
\label{QMA} V_{\rm MAG}&=&
\frac{1}{2\kappa}\,\left[-a_0\,R^{\alpha\beta}\wedge\eta_{\alpha\beta}
  -2\lambda_{0}\,\eta+T^\alpha\wedge{}^*\!\left(\sum_{I=1}^{3}a_{I}\,^{(I)}
    T_\alpha\right)\right.\nonumber\\ &+&\left.
  2\left(\sum_{I=2}^{4}c_{I}\,^{(I)}Q_{\alpha\beta}\right)
  \wedge\vartheta^\alpha\wedge{}^*\!\, T^\beta + Q_{\alpha\beta}
  \wedge{}^*\!\left(\sum_{I=1}^{4}b_{I}\,^{(I)}Q^{\alpha\beta}\right)\right.
\nonumber \\&+&
b_5\bigg.\left(^{(3)}Q_{\alpha\gamma}\wedge\vartheta^\alpha\right)\wedge
{}^*\!\left(^{(4)}Q^{\beta\gamma}\wedge\vartheta_\beta \right)\bigg]
\nonumber\\&- &\frac{1}{2\rho}\,R^{\alpha\beta} \wedge{}^*\!
\left(\sum_{I=1}^{6}w_{I}\,^{(I)}W_{\alpha\beta}
  +w_7\,\vartheta_\alpha\wedge(e_\gamma\rfloor
  ^{(5)}W^\gamma{}_{\beta} ) \nonumber\right.\\&+& \left.
  \sum_{I=1}^{5}{z}_{I}\,^{(I)}Z_{\alpha\beta}+z_6\,\vartheta_\gamma\wedge
  (e_\alpha\rfloor ^{(2)}Z^\gamma{}_{\beta}
  )+\sum_{I=7}^{9}z_I\,\vartheta_\alpha\wedge(e_\gamma\rfloor
  ^{(I-4)}Z^\gamma{}_{\beta} )\right)\quad\,.
\end{eqnarray}
Here $\kappa$ is the dimensionful (weak) gravitational constant,
$\lambda_{0}$ the ``bare'' cosmological constant, and the
dimensionless $\rho$ is the strong gravity coupling constant. The
constants $ a_0, \ldots a_3$, $b_1, \ldots b_5$, $c_2, c_3,c_4$, $w_1,
\ldots w_7$, $z_1, \ldots z_9$ are dimensionless and should be of
order unity. Note the nontrivial formula for the Hilbert-Einstein type
of Lagrangian
\begin{equation}\label{RW6}
R_{\a\b}\wedge\eta^{\a\b}=\,^{(6)}W_{\a\b}\wedge\eta^{\a\b}\,.
\end{equation}
Because of the irreducibility, we have
\begin{eqnarray}
  T^2&\sim& T^\alpha\wedge{}^\star\left(a_{1}\,^{(1)}
    T_\alpha+a_{2}\,^{(2)} T_\alpha+a_{3}\,^{(3)}
    T_\alpha\right)\nonumber\\ &=& a_1 \,^\star{}^{(1)} T_\alpha\wedge
  {}^{(1)} T_\alpha+ a_2 \,^\star{}^{(2)} T_\alpha\wedge {}^{(2)}
  T_\alpha+ a_3 \,^\star{}^{(3)} T_\alpha\wedge {}^{(3)} T_\alpha\,,
\end{eqnarray}
and similar formulas for the pure square pieces of $Q_{\a\b}$,
$W_{\a\b}$, and $Z_{\a\b}$. In the curvature square terms in
(\ref{QMA}) we introduced the irreducible pieces of the antisymmetric
part $W_{\alpha\beta}:= R_{[\alpha\beta]}$ (rotation part) and the
symmetric part $Z_{\alpha\beta}:= R_{(\alpha\beta)}$ (strain part) of
the curvature 2--form $R_{\alpha\beta}$.  In $Z_{\alpha\beta}$, we
have the purely {\em post}--Riemannian part of the curvature. Note the
peculiar cross terms with $c_I$ and $b_5$, and the {\em exotic\/}
terms with $w_7$ and $z_6,z_7,z_8,z_9$.

In the component formalism, Esser \cite{Esser} has carefully
enumerated all different pieces of the quadratic MAG Lagrangian; for
the corresponding nonmetricity and torsion pieces, see also Duan et
al.\ \cite{Duan}.  Accordingly, Eq.(\ref{QMA}) represents the most
general quadratic parity--conserving MAG--Lagrangian. All previously
published quadratic parity--conserving Lagrangians are subcases of
(\ref{QMA}).  Hence (\ref{QMA}) is a safe starting point for our
future considerations.

%%%%%%%%%%%%%%%%%%%%%%%%%%%%%%%%%%%%%%%%%%%%%%%%%%%%%%%%%%%%%%%%%%%
\section{The 2-form $\ra$ and the first Bianchi
  identity}\label{firstBia}
%%%%%%%%%%%%%%%%%%%%%%%%%%%%%%%%%%%%%%%%%%%%%%%%%%%%%%%%%%%%%%%%%%%

The first Bianchi identity  $B^\a\equiv 0$, with the 3-form
\begin{equation}\label{firstbia}
  B^\alpha := DT^\alpha - R_\beta{}^\alpha \wedge \vta^\beta\,,
\end{equation}
interrelates torsion and curvature.  The irreducible decomposition of
the first Bianchi identity reads (see \cite{PRs})
\begin{eqnarray}
{}^{(2)}B^\alpha
&=&
\frac{1}{n-2} \, (e_\beta \rfloor B^\beta) \wedge \vta^\alpha \,,\\
{}^{(3)}B^\alpha
&=&
\frac{1}{4} \, e_\alpha \rfloor (\vta^\beta \wedge B_\beta) \,, \\
{}^{(1)}B^\alpha
&=&
B^\alpha - {}^{(2)}B^\alpha - {}^{(3)}B^\alpha \,.
\end{eqnarray}

In the last term of (\ref{firstbia}), because of the contraction with
the coframe, some of the irreducible components of the curvature drop
out, see \cite{PRs}, Eqs.(B.4.15) and (B.4.29):
\begin{equation}
R_\beta{}^\alpha \wedge \vta^\beta
=
\left(   {}^{(2)}W_\beta{}^\alpha
       + {}^{(3)}W_\beta{}^\alpha
       + {}^{(5)}W_\beta{}^\alpha
       + {}^{(2)}Z_\beta{}^\alpha
       + {}^{(3)}Z_\beta{}^\alpha
       + {}^{(4)}Z_\beta{}^\alpha
\right) \wedge \vta^\beta \,.
\end{equation}
We contract this piece by the frame:
\begin{eqnarray}
e_\alpha \rfloor \left( R_\beta{}^\alpha\wedge \vta^\beta\right)
&=&
e_\alpha \rfloor \left( {}^{(5)}W_\beta{}^\alpha\wedge \vta^\beta\right)
+e_\alpha \rfloor \left( {}^{(3)}Z_\beta{}^\alpha\wedge \vta^\beta\right)
+e_\alpha \rfloor \left( {}^{(5)}Z_\beta{}^\alpha\wedge \vta^\beta\right)
\nonumber\\&=&
- \ra - (n-2) \, \hat\Delta +\frac{n-2}{2} \, dQ \,.\label{dQ}
\end{eqnarray}
We introduced as abbreviation the
antisymmetric rotational Ricci 2-form for $W_{\a\b}$ by
\begin{equation}\label{frag}
  \ra := \vta^\alpha \wedge (e_\beta \rfloor
  \,^{(5)}W_{\alpha}{}^{\beta})= \vta^\alpha \wedge (e_\beta \rfloor
  W_{\alpha}{}^{\beta}) \,,
\end{equation}
see \cite{PRs}, Eqs.(B.4.8) and (B.4.11). $\ra$ has $n(n-1)/2$
independent components. Note that there is a subtlety involved here.
The (premetric) Ricci 1-form is usually defined in terms of the {\it
  total\/} curvature according to ${\rm Ric}_{\a}:=e_\b \rfloor
R_\a{}^\b$. Here, however, our definition refers only to the
rotational curvature $W_{\a\b}$, see (\ref{Wdeco}).  Similarly we
obtain
\begin{equation}\label{tildeX}
  \vta^\alpha \wedge \left(R_{\b\a}\wedge\vta^\beta\right) =
  \vta^\alpha \wedge
  \left({}^{(3)}W_{\beta\a}\wedge\vta^\beta\right) =: \hat{X}
  \,.
\end{equation}
Before we substitute (\ref{dQ}) and (\ref{tildeX}) into the first
Bianchi identity, we should first compute the corresponding torsion
pieces:
\begin{eqnarray}
e_\alpha \rfloor DT^\alpha
&=&
e_\alpha \rfloor \left[ D({}^{(1)}T^\alpha + {}^{(3)}T^\alpha)\right]
+\frac{1}{n-1} \, \left({}^{(1)}T^\alpha + {}^{(3)}T^\alpha\right) \,
e_\alpha \rfloor T \nonumber \\ &&
- \frac{n-2}{n-1} \, dT \,,\\
\vta^\alpha \wedge DT_\alpha
&=&
-D(\vta^\alpha \wedge {}^{(3)}T_\alpha )
+ T^\alpha \wedge T_\alpha \,.
\end{eqnarray}
In this way, we find for the 3-form $B^\a$ eventually,
\begin{eqnarray}
  e_\beta \rfloor B^\beta &=& e_\alpha \rfloor \left[
    D({}^{(1)}T^\alpha + {}^{(3)}T^\alpha)\right] +\frac{1}{n-1} \,
  ({}^{(1)}T^\alpha + {}^{(3)}T^\alpha) \, e_\alpha \rfloor T
  \nonumber\\ &&-\frac{n-2}{n-1} \, dT +\ra + (n-2) \, \hat{\Delta}
  -\frac{n-2}{2} \, dQ =0\,,\\ \vta^\beta \wedge B_\beta
  &=& - D(\vta^\alpha \wedge {}^{(3)}T_\alpha ) + T^\alpha \wedge
  T_\alpha -\hat{X} = 0\,.
\end{eqnarray}
Provided the torsion possesses {\it only its trace piece}, that is
${}^{(1)}T^\alpha={}^{(3)}T^\alpha=0$, and furthermore
${}^{(2)}Q_{\alpha\beta}=0$, see the constraints (\ref{circ}), we have
$\hat{\Delta}=d\Lambda/[2(n-1)]$ and
\begin{equation}\label{integrab}
\ra - \frac{n-2}{2(n-1)} \, d \left[2T + (n-1) \, Q -\Lambda\right]
=0 \,.
\end{equation}
Thus, in $n=4$, with the constraints (\ref{circ}) fulfilled and for
$\ra=0$, the first Bianchi identity yields
\begin{equation}\label{letzte}
  d\,(3Q + 2T -{\Lambda})=0 \,.
\end{equation}
Accordingly, under these conditions, the Weyl, the torsion, and the
shear 1-forms are algebraically related to each other.

%%%%%%%%%%%%%%%%%%%%%%%%%%%%%%%%%%%%%%%%%%%%%%%%%%%%%%%%%%%%%%%
\section{Curvature of the spherically symmetric aether solution
of Sec.\ref{simplesolution}} \label{cursol1}
%%%%%%%%%%%%%%%%%%%%%%%%%%%%%%%%%%%%%%%%%%%%%%%%%%%%%%%%%%%%%%%

With the help of our Reduce-Excalc computer algebra programs, we
calculate the rotational and the strain curvature, respectively (the
diamonds $\diamond$ and the bullets $\bullet$ denote those matrix
elements that are already known because of the antisymmetry or the
symmetry of the matrix involved):
\begin{eqnarray}\label{F1}
{}^{(1)}W_{\alpha\beta} 
&=&
\left(\frac{m}{r^3} -\frac{({\ell}_0+{\ell}_1)(4{\ell}_1-{\ell}_0)}{96r^4 \, 
e^{2\mu(r)}} \right)
\, \left(\begin{array}{cccc}
0 & 2 \vta^{01} & -\vta^{02} & -\vta^{03} \\
\diamond & 0 & \vta^{12} & \vta^{13} \\
\diamond & \diamond & 0 & -2 \vta^{23} \\
\diamond & \diamond & \diamond & 0
\end{array} \right) \qquad\hbox{({\tt weyl})}\,, \\
{}^{(2)}W_{\alpha\beta} 
&=&
0 \qquad\hbox{({\tt paircom} = 0)} \,,\\
{}^{(3)}W_{\alpha\beta}
&=&
0\qquad\hbox{({\tt pscalar} = 0)} \,,\\
{}^{(4)}W_{\alpha\beta} 
&=&
- \vta_{[\alpha}\wedge\phi_{\beta]}\qquad\hbox{({\tt ricsymf})}  \,,\\
\phi_0 \nonumber
&=& 
 \frac{(\ell_0+7\ell_1)(\ell_0-\ell_1)\,\vta^0 - 4  \ell_0 \ell_1 \,
\vta^1}{32r^4 e^{2\mu(r)}}\,, \\
\phi_1 \nonumber
&=& -\frac{4  \ell_0\ell_1 \,\vta^0 + 
(\ell_0-5\ell_1)(\ell_0-\ell_1)\,\vta^1}{32 r^ 4e^{2\mu(r)}}\,,\\
\phi_2  \nonumber
&=& 
\frac{(\ell_0+\ell_1)(\ell_0-\ell_1)}{32 r^ 4e^{2\mu(r)}} \; \vta^2 \,,\\
\phi_3 \nonumber
&=&
\frac{(\ell_0+\ell_1)(\ell_0-\ell_1)}{32 r^ 4e^{2\mu(r)}} \; \vta^3 \,,\\
{}^{(5)}W_{\alpha\beta} 
&=& 
0 \qquad\hbox{({\tt ricanti} = 0)}  \,,\\
{}^{(6)}W_{\alpha\beta} \label{F6}
&=&
-\frac{1}{12} \, W \, \vta_{\alpha\beta} \,, \qquad
W = 4 \lambda_0 + \frac{(\ell_0+5\ell_1)(\ell_0+\ell_1)}{8r^4 e^{2\mu(r)}}
 \qquad\hbox{({\tt scalar})}\,.\\ &&\nonumber\\
{}^{(1)}Z_{\alpha\beta} 
&=&\hspace{-3pt}
\frac{1}{48r^4 e^{2\mu(r)}}\!  \left(
\begin{array}{cccc}
  -4A_1\, \vta^{01} & -2A_2 \, \vta^{01} & A_2\, \vta^{02} + 2 A_1 \,
  \vta^{12} & A_2 \vta^{03} + 2A_1\,\vta^{13} \vspace{4pt} \\ \bullet
  & 4 A_3\, \vta^{01}& -2A_3\, \vta^{02} +A_2 \, \vta^{12} & -2A_3 \,
  \vta^{03} +A_2\,\vta^{13} \vspace{4pt}\\ \bullet & \bullet & 24
  \ell_1 \, r \, e^{2\mu(r)} \, \vta^{01} & 0 \vspace{4pt}\\ \bullet &
  \bullet & \bullet & 24\ell_1 \,r\,e^{2\mu(r)} \, \vta^{01}
\end{array} \right)\!,\\
&&
A_1 = \ell_1 (4\lambda_0 r^3 + 15m -9r) \,, \qquad \nonumber
A_2 = \ell_0 (2\lambda_0 r^3 + 3m - 3r) + 9 \ell_1 (3m -r) \,, \\
&&
A_3 = 3 \ell_1 (3m-r) \,, \nonumber \\
{}^{(2)}Z_{\alpha\beta} 
&=&
0 \,, \\
{}^{(3)}Z_{\alpha\beta} 
&=& 0 \,, \\
{}^{(4)}Z_{\alpha\beta} &=& -\frac{\ell_1}{2r^3} \, g_{\alpha\beta} \,
\vta^{01} \qquad\hbox{({\tt dilcurv})} \,,\\
{}^{(5)}Z_{\alpha\beta} &=&\frac{1}{2} \, \vta_{(\alpha} \wedge \Xi_{\beta)}
\,, \\
\Xi_0 &=& -\frac{(\lambda_0 r^3-3m)(\ell_0+\ell_1)\,\vta^0 - 2(3m-r)\ell_1\,
\vta^1}{4r^4e^{2\mu(r)}} \,,\nonumber\\
\Xi_1 &=& -\frac{ \left[3(\ell_0+9\ell_1)m -6 (\ell_0+3\ell_1)r
+(5\ell_0+9\ell_1)\lambda_0 \, r^3\right] \, \vta^1
             -6(3m-r)\ell_1 \,\vta^0}{12r^4e^{2\mu(r)} } \,,\nonumber\\
\Xi_2 &=&  -\frac{\ell_0+3\ell_1}{4r^3} \; \vta^2 \,,\nonumber\\
\Xi_2 &=&  -\frac{\ell_0+3\ell_1}{4r^3} \; \vta^3 \,.\nonumber
\end{eqnarray}

\centerline{================}

\end{document}